\documentclass[12pt]{article}
\usepackage[margin=1in]{geometry}

\usepackage{hyperref}
\usepackage{color}

\usepackage[english]{babel}
\usepackage{amsmath,amsthm}
\usepackage{amssymb}
\usepackage{amsfonts}

\usepackage{algorithm}  
\usepackage{algorithmicx}  
\usepackage{algpseudocode}

\usepackage{CJKutf8}
\usepackage{graphicx}
\usepackage{pgf,tikz}
\usetikzlibrary{shapes,arrows,automata}

\newtheorem{thm}{Theorem}[section]

\newtheorem{lem}[thm]{Lemma}
\newtheorem{conj}[thm]{Conjecture}

\theoremstyle{definition}
\newtheorem{defn}[thm]{Definition}
\newtheorem{rem}[thm]{Remark}

\newtheorem{fact}[thm]{Fact}
\numberwithin{equation}{section}

\newcommand{\tKW}{\textnormal{KW}}
\newcommand{\tU}{\textnormal{U}}

\newcommand{\defeq}{\stackrel{\text{def}}{=}}

\begin{document}
\title{An Improved Composition Theorem of a Universal Relation and Most Functions via Effective Restriction}%
\author{Hao Wu\thanks{College of Information Engineering, Shanghai Maritime University, Shanghai, China. My email is \texttt{haowu@shmtu.edu.cn}, you can also reach me via \texttt{wealk@outlook.com}.}}%
\maketitle
\begin{abstract}
One of the major open problems in complexity theory is to demonstrate an explicit function which requires super logarithmic depth, to tackle this problem Karchmer, Raz and Wigderson proposed the KRW conjecture about composition of two functions. While this conjecture seems out of our current reach, some relaxed conjectures are suggested to be the stepping stone to the original one. One important kind of relaxed forms is  composition about universal relation. We already have strong lower bounds for composition of two universal relations as well as composition of a function and a universal relation. The final jigsaw to complete our understanding of composition about universal relation is the composition of a universal relation  and a function. Recently, Ivan Mihajlin and Alexander Smal proved a composition theorem of a universal relation and some function via so called xor composition, that is there exists some function $f:\{0,1\}^n \rightarrow \{0,1\}$ such that $\textsf{CC}(\textnormal{U}_n \diamond \text{KW}_f) \geq 1.5n-o(n)$ where $\textsf{CC}$ denotes the communication complexity of the problem.

In this paper, we significantly improve their result and present an asymptotically tight and much more general composition theorem of a universal relation and most  functions, that  is for most functions $f:\{0,1\}^n \rightarrow \{0,1\}$ we have $\textsf{CC}(\textnormal{U}_m \diamond \text{KW}_f) \geq m+ n -O(\sqrt{m})$ when $m=\omega(\log^2 n),n =\omega(\sqrt{m})$. This is done by  a direct proof of composition theorem of a universal relation and a multiplexor in the partially half-duplex model avoiding the xor composition. And the proof works even when the multiplexor only contains a few functions. One crucial ingredient in our proof involves a combinatorial problem of constructing a tree of many leaves and every leaf contains a non-overlapping set of functions. For each leaf, there is a set of inputs such that every function in the leaf takes the same value, that is all functions are restricted. We show how to choose a set of good inputs to effectively restrict these functions to force that the number of  functions  in each leaf is as small as possible  while maintaining the total number of functions in all leaves. This results in a large number of leaves.
\end{abstract}

\newpage
\tableofcontents
\newpage

\section{Introduction}
One of the major open problems in complexity theory is to demonstrate an explicit function which requires super logarithmic depth, a.k.a, the $\mathbf{P}$ versus $\mathbf{NC^1}$ problem.  The current best depth lower bound \cite{DBLP:journals/siamcomp/Hastad98,DBLP:conf/focs/Tal14} is $(3-o(1))\cdot \log n$, and we still don't even know how to obtain  a lower bound strictly larger than $3\log n$. One promising approach to tackle this problem was suggested by  Karchmer, Raz and Wigderson \cite{DBLP:journals/cc/KarchmerRW95}, they proposed that we should understand the complexity of  (block)-composition of Boolean functions. Given two  functions $f: \{0,1\}^m \rightarrow  \{0,1\}$, $g: \{0,1\}^n \rightarrow  \{0,1\}$, we define their composite function $f\diamond g:(\{0,1\}^n)^m \rightarrow  \{0,1\}$ as:
$
f \diamond g\left(x_{1}, \ldots, x_{m}\right)=f\left(g\left(x_{1}\right), \ldots, g\left(x_{m}\right)\right).
$
Given any Boolean function $f$, we denote the depth complexity of $f$ by $\mathsf{D}(f)$, that is  the minimal depth of a circuit of AND, OR and NOT gates of fan-in $2$ that computes $f$.  And it is easy to see the depth complexity of  $f \diamond g$  is upper-bounded by $\mathsf{D}(f)+\mathsf{D}(g)$ and it is natural to ask whether the depth complexity of  $f \diamond g$  is far from this upper bound. Karchmer, Raz and Wigderson \cite{DBLP:journals/cc/KarchmerRW95} conjectured that the depth complexity of  $f \diamond g$  is not far from its upper bound:
\begin{conj}Given two arbitrary non-constant Boolean functions $f:\{0,1\}^{m} \rightarrow\{0,1\}$ and $g:\{0,1\}^{n} \rightarrow\{0,1\}$, then
	$
	\mathsf{D}(f \diamond g) \approx \mathsf{D}(f)+\mathsf{D}(g).
	$
\end{conj}
The merit of this conjecture is, if it is proved and the ``approximate equality'' is instantiated with proper parameters, then by an argument of  iterative composition \cite{DBLP:journals/cc/KarchmerRW95}, we will obtain an explicit function with super-logarithmic depth, which separates $\mathbf{P} $ from $\mathbf{NC^1}$. The hope to resolve this conjecture lies in a deep and elegant connection between circuit complexity and communication complexity which is captured by the concept of Karchmer-Wigderson  relations \cite{DBLP:journals/siamdm/KarchmerW90}. Given a Boolean function $f:\{0,1\}^{n} \rightarrow$ $\{0,1\},$ the Karchmer-Wigderson relation (KW relation  for short) of function $f$, denoted by $\tKW_{f}$, is the following communication problem:  Alice gets an input $x \in f^{-1}(1),$ and Bob gets an input $y \in f^{-1}(0)$. The goal of Alice and Bob is to find a coordinate $i \in [n]$ such that $x_{i} \neq y_{i}$. Note that since $x\neq y$, there always exists at least one such coordinate.

The key observation by Karchmer and Wigderson \cite{DBLP:journals/siamdm/KarchmerW90} is that the \emph{deterministic} communication complexity of $\tKW_{f}$ is exactly equal to $\mathsf{D}(f)$. This allows us to view the original KRW conjecture from the $\mathrm{KW}$ relation perspective. Let $ f:\{0,1\}^{m} \rightarrow\{0,1\}$ and $g:\{0,1\}^{n} \rightarrow$ \{0,1\} be non-constant functions.  In the KW relation $K W_{f \diamond g}$, the inputs to Alice and Bob are viewed as two $m \times n$ Boolean matrices $X,Y$. Alice gets $X \in(f \diamond g)^{-1}(1)$ and Bob gets $Y \in(f \diamond g)^{-1}(0)$, their task is to find an entry $(i, j)$ such that $X_{i, j} \neq Y_{i, j}$. Moreover, it is convenient to write $\tKW_{f \diamond g}$ as $\tKW_{f }\diamond \tKW_{g}$, indicating that these KW relations could be more general KW relation rather than KW relation of functions, now we can rephrase KRW conjecture in terms of communication complexity:
\begin{conj}Given two arbitrary non-constant Boolean functions $f:\{0,1\}^{m} \rightarrow\{0,1\}$ and $g:\{0,1\}^{n} \rightarrow\{0,1\}$, then
	$
	\mathsf{CC}\left(\tKW_{f }\diamond \tKW_{g}\right) \approx \mathsf{CC}\left(\tKW_{f}\right)+\mathsf{CC}\left(\tKW_{g}\right),
	$
	where $\mathsf{CC}$ means the deterministic communication complexity of a KW relation.
\end{conj}
Current successes towards KRW conjecture are all restricted cases. There are composition theorems when the inner function $g$ satisfies certain property, for example when the inner function is the parity function \cite{DBLP:journals/siamcomp/Hastad98,DBLP:conf/focs/Tal14,DBLP:journals/cc/DinurM18} and when  the inner functions are with a  tight unweighted quantum
adversary bound \cite{DBLP:conf/innovations/FilmusMT21}. There are composition theorems where the composition itself is restricted such as monotone composition, semi-monotone composition \cite{DBLP:conf/focs/RezendeMNPR20} and strong composition \cite{DBLP:journals/eccc/Meir23}. There are also some variants \cite{DBLP:journals/cc/EdmondsIRS01,DBLP:journals/cc/Meir20,DBLP:conf/coco/MihajlinS21} of original conjecture with the similar effect to the $\mathbf{P}$ versus $\mathbf{NC^1}$ problem, but we don't know how to prove them either. Maybe to prove the general form of KRW conjecture is out of our reach now.
Edmonds, Impagliazzo, Rudich and Sgall  \cite{DBLP:journals/cc/EdmondsIRS01} suggested we should consider relaxed form of KRW conjecture and hope that any progresses of these relaxed compositions involve ideas and techniques  which  will be useful to attack the original KRW conjecture. One choice is to relax the KW relation of function to the universal relation. In  the universal relation $\operatorname{U}_{n}$, Alice and Bob get two distinct strings $x, y \in\{0,1\}^{n}$,  their task is to find a coordinate $i$ such that $x_i \neq y_i$. It is perhaps a necessary starting point for us to study composition of KW relations.

The first challenge is to prove lower bound for composition of two  universal relations $\tU_m\diamond \tU_n$, this  was met by 
\cite{DBLP:journals/cc/EdmondsIRS01,HW93}. The next step is to understand the composition of a function and a universal relation $\tKW_f\diamond \tU_n$. 
Gavinsky, Meir, Weinstein and Wigderson 
\cite{DBLP:journals/siamcomp/GavinskyMWW17} showed a lower bound with a small additive loss, then Koroth and Meir \cite{Koroth2018} improved their result and provided an essential optimal lower bound for $\tKW_f\diamond \tU_n$. After that, the final jigsaw to complete our understanding of composition  about universal relation is composition of a universal relation and a function.
Let $f:\{0,1\}^{n} \rightarrow \{0,1\}$  be a function. Given any Boolean matrix $X\in \{0,1\}^{m \times n}$, define $f(X)=(f(X_1),\ldots,f(X_m))$. In  KW relation $\tU_m \diamond \tKW_f$ , Alice gets a Boolean matrix $X\in \{0,1\}^{m \times n}$,  Bob gets a Boolean matrix $Y\in \{0,1\}^{m \times n}$,  their goal is to find an entry $(i, j)$ such that $X_{i, j} \neq Y_{i, j}$. If  $f(X)=f(Y)$, they can also output $\bot$. It is natural to make following conjecture \cite{DBLP:journals/siamcomp/GavinskyMWW17,DBLP:journals/cc/DinurM18}.
\begin{conj}\label{1.3}Given a universal relation $\operatorname{U}_{m}$ and a function $ f:\{0,1\}^{n} \rightarrow\{0,1\}$, then
	$
	\mathsf{CC}\left(\operatorname{U}_{m}\diamond \tKW_{ f}\right) \approx m+\mathsf{CC}\left(\tKW_{f}\right).
	$
\end{conj}
Ivan Mihajlin and Alexander Smal \cite{DBLP:conf/coco/MihajlinS21} took a big step towards Conjecture \ref{1.3} and proved a composition theorem of a universal relation and some function via so called xor composition, that is there exists some function $f:\{0,1\}^n \rightarrow \{0,1\}$ such that $\textsf{CC}(\textnormal{U}_n \diamond \text{KW}_f) \geq 1.5n-o(n)$. But their result is not tight and works only for some function when $m\simeq n$, they asked whether the success of \cite{DBLP:journals/siamcomp/GavinskyMWW17,Koroth2018} can be achieved in the case of $\textnormal{U}_n \diamond \text{KW}_f$. Thus comparing to the optimal lower bound in the case of $\tKW_{ f}\diamond \operatorname{U}_{n}$, following conjecture should not be too ambitious.
\begin{conj}\label{1.4}Given a universal relation $\operatorname{U}_{m}$ and a function $ f:\{0,1\}^{n} \rightarrow\{0,1\}$ with $m,n$  in proper range, then
	$
	\mathsf{CC}\left(\operatorname{U}_{m}\diamond  \tKW_{f}\right) = m+\mathsf{CC}\left(\tKW_{f}\right)-o(\min\{m,\mathsf{CC}(\tKW_f)\}).
	$
\end{conj}
In this paper, we make progress towards Conjecture \ref{1.4} and show it is almost true.
\subsection{Our results}
Our main result is for most  functions $f:\{0,1\}^n \rightarrow \{0,1\}$, $\mathsf{CC}(\textnormal{U}_m\diamond \textnormal{KW}_f) \geq m+n-O(\sqrt{m})$.
\begin{thm}\label{1.5}Let  $m=\omega(\log^2 n),n =\omega(\sqrt{m})$, if we pick a  function $f:\{0,1\}^n \rightarrow \{0,1\}$ randomly,  the probability of $\mathsf{CC}(\textnormal{U}_m \diamond \textnormal{KW}_f) \geq m+n-O(\sqrt{m})$ is $1-o(1)$.
\end{thm}
This result follows from a composition theorem of a universal relation and a multiplexor in the partially half-duplex model. And it works even when the multiplexor only contains a few functions.	Let $\mathcal{F}$ be a set of functions $f : \{0,1\}^n \rightarrow \{0,1\}$. In  KW relation $\tU_m\diamond \textnormal{MUX}_\mathcal{F}$, Alice gets a function $f\in \mathcal{F}$ and a Boolean matrix $X\in \{0,1\}^{m \times n}$,  Bob gets a function $g \in \mathcal{F}$ and a Boolean matrix $Y\in \{0,1\}^{m \times n}$,  their goal is to find an entry $(i, j)$ such that $X_{i, j} \neq Y_{i, j}$. If $f \neq g$ or $f(X)=g (Y)$, they can also output $\bot$.
\begin{thm}\label{1.6}
Let  $m=\omega(\log^2 n),n =\omega(\sqrt{m})$, $\epsilon =\frac{\sqrt{m}}{n}$. Let  $\mathcal{F}$ be a set of functions $f:\{0,1\}^n \rightarrow \{0,1\}$ such that $|\mathcal{F}|\geq 2^{2^n-2^{(1-\epsilon)n}}$ we have
	$\mathsf{CC^{phd}}(\textnormal{U}_m \diamond \textnormal{MUX}_\mathcal{F}) \geq m+n-O(\sqrt{m})$
where $\mathsf{CC^{phd}}$ denotes the communication complexity in partially half-duplex model. 
\end{thm}
\paragraph{Comparison with related works.}Comparing to the result of Ivan Mihajlin and Alexander Smal, our result is asymptotically tight and much more general. More importantly, we give a direct proof without using the xor composition. We also note recently Meir \cite{DBLP:journals/eccc/Meir23} proved  a result about `strong' composition of a function and a multiplexor, but so-called strong composition is a restricted form of the standard composition while our result is about standard composition, thus our result is incomparable to Meir's.
\paragraph{Our approach.}Here we give a simplified description of our proof of Theorem \ref{1.6}. For convenience, assume $n\geq m$ and ignore the difference between standard communication model and the partially half-duplex model.
We can prove it via a two-stage argument similar to that in \cite{DBLP:conf/coco/MihajlinS21}, that is after the protocol has spent approximate $m$ bits, we are able to extract a set $\mathcal{H}$ of size almost $2^{2^n}$ from the residual problem,  then use this set and  the protocol to solve the non-equality problem over $\mathcal{H}$ non-deterministically, thus the protocol will require another approximate $n$ bits. Now we give more details. Let $c$ be an integer which depends on $m,n,\epsilon$. Let $t=c+4$, $s=m-t-1$, $\mathcal{X}$ be the set $\{0,1\}^{m\times n}$, $\mathcal{D}=\{((f,X),(f,X))\mid f\in \mathcal{F},X\in \mathcal{X}\}$ and $d$ be the depth of the protocol. 
\begin{itemize}
	\item In the first stage, there is a (partial) transcript $\tau\in \{0,1\}^s$ and a subset of inputs  $\mathcal{D}^\prime \subseteq \mathcal{D}$ such that every input in $\mathcal{D}^\prime$ is consistent with $\tau$. Intuitively, after spending the $s$ bits in the transcript $\tau$, the residual protocol still has to solve all inputs from the set $\mathcal{D}^\prime$. Furthermore, there is a set $\mathcal{S}\subseteq \mathcal{F} \times \mathcal{X}$ such that $\{((f,X),(f,X))\mid (f,X)\in \mathcal{S} \} \subseteq \mathcal{D}^\prime$ and 
	 \begin{itemize}
	 	\item $|\mathcal{S}|   \geq 2^{t-m}\cdot|\mathcal{F}| \cdot |\mathcal{X}|$.
	 	\item  Let $\mathcal{U}_\mathcal{S} =\{f\mid(f,X)\in \mathcal{S}\}$,
	 	for every $f\in \mathcal{U}_\mathcal{S}$,   $|\mathcal{X}_{\mathcal{S},f}| \geq 2^{t-m}\cdot |\mathcal{X}|$.
	 \end{itemize}
     Eventually we can extract a subset  $\mathcal{H}\subseteq \mathcal{U}_\mathcal{S}$ of size at least $2^{2^{(1-\epsilon)n}}$  such that for all distinct $f, g \in \mathcal{H}$, there exists an $ X:
 (f,  X), (g,  X)\in \mathcal{S}$, and $f(X) \neq  g(X)$. 
 \item In the second stage, recall that the depth of residual protocol is at most $d-s$ and  by the rectangle property it must correctly  solve every input from set $\{(f,X),(g,  X)\mid (f,X),(g,X)\in \mathcal{S},f, g \in \mathcal{H}\}$. We can  leverage this fact to non-deterministically  solve the non-equality problem over $\mathcal{H}$ with a witness of size $d-s+O(\sqrt{m})$.
\end{itemize}
Since the nondeterministic complexity of the non-equality problem over $\mathcal{H}$ is at least $\log\log\mathcal{H} $, we have $d-s+O(\sqrt{m})\geq\log\log\mathcal{H} $, that is the depth $d\geq m+n-O(\sqrt{m})$.

Let's take a glimpse at how to effectively extract the set $\mathcal{H}$, see more details in Lemma~\ref{3.4}. The extraction involves a combinatorial problem of constructing a tree and every leaf of the tree contains a non-overlapping set of functions, then the set $\mathcal{H}$ takes one function from each leaf. The tree is constructed recursively. Each node $z$ in the tree is associated with a set $\mathcal{Z} \subseteq \mathcal{S}$, let $\mathcal{U}_\mathcal{Z} =\{f \mid (f,X)\in \mathcal{Z}\}$, then every node in the same depth contains a non-overlapping set $\mathcal{U}_\mathcal{Z}$ of functions. Assume $z$ is at depth $d$, from root to node $z$, its ancestors are $z_0,z_1,\ldots,z_{d-1}$. For every $i\in \{0,1,\ldots,d-1\}$, $z_i$ is labeled with  $X_{z_i}$, treat every $X_{z_i}$ as a set of its distinct rows, we define $\Psi(z) = \bigcup_{i=0}^{d-1} X_{z_i}$. Then given inputs from $\Psi(z)$, every $f \in \mathcal{U}_\mathcal{Z}$ takes the same value  thus restricted in $\{0,1\}^m \setminus \Psi(z)$. Now it's turn to choose a good $X_z$ for node $z$ to restrict functions  in the children of $z$ as much as possible, meanwhile maintaining the total number of functions in all its children. Fortunately, we can choose a good $X_z$ in each step downward, such that  the number of functions in each child decreases by a factor of  (at least) $2^{m-c}$ while the total number of functions in all its children decreases  by a smaller (average) factor of (at most) $2^{m+3-t}$. The parameters are carefully chosen to make sure that $t=c+4$, and  finally, at depth $h=2^{\lceil(1-\epsilon)n\rceil}$, the total number of functions in all leaves is $(2^{-(m+3-t)}/2^{-(m-c)})^h=2^{(m-c-(m+3-t))h}=2^h$  times bigger than the number of  functions  in each leaf, thus we obtain a set $\mathcal{H}$ of size at least $2^{2^{(1-\epsilon)n}}$.
We have omitted some technicalities in the full proof as follows.
\begin{itemize}
	\item When $n \ll m$, in the first stage, we can only obtain  a short transcript $\tau$ such that $|\tau|\approx n\ll m$, thus  single shot of two-stage augment is not sufficient. Nevertheless, we can use the two-stage argument multiple times to boost the complexity up until it's done. See the discussion at the beginning of Section \ref{s3}.
	\item The second problem is the difference between the standard communication model and the partially half-duplex model, and the argument has to be tuned to be compatible with the partially half-duplex model. Nonetheless, this problem can be overcome in a similar way like that in \cite{DBLP:conf/coco/MihajlinS21}, see more details in Section \ref{s2.3} and Lemma \ref{3.2}.
\end{itemize}
\subsection{Organization of the rest of the paper}
The rest of the paper is organized as follows. In Section \ref{s2}, we provide necessary preliminaries. It is highly recommended not to skip Section \ref{s2.2} and \ref{s2.3}, particularly, we explain how we avoid xor composition in Section \ref{s2.2}. In Section \ref{s3}, we prove Theorem \ref{1.6}, a composition theorem of a universal relation and a multiplexor in  the model of partially half-duplex communication with adversary. In Section \ref{s4}, we prove Theorem \ref{1.5}, a composition theorem of a universal relation and most functions in  the standard model of communication. In Section~\ref{s5}, we make some discussion and point out some future directions.

\section{Preliminaries and Notations}\label{s2}
In this section, we provide some basic notations, definitions and facts. Let $\mathbb{N}^+$ be the set of positive natural numbers, for any $n \in \mathbb{N}^+$, we denote by $[n]$ the set $\{1, \ldots, n\}$. Let $x\in \{0,1\}^n$ be a Boolean string, we denote the $i$-th bit of $x$ by $x_i$. Let $X\in  \{0,1\}^{m \times n}$ be an $m\times n$ Boolean matrix, we denote the $i$-th row of $X$ by $X_i$ and the entry at $(i,j)$ by $X_{i,j}$.
\subsection{Communication complexity}
We assume the readers are familiar with the basic knowledge of communication complexity, a more detailed introduction to communication complexity can be found in textbooks such as \cite{DBLP:books/daglib/0011756,rao_yehudayoff_2020}. 

\begin{defn}[Two party communication problems]In a two-party communication problem ${S} \subseteq ({X} \times {Y}) \times Z$, there are two involved players, Alice and Bob, who need to solve following task: 
	Alice is given an input $x \in X$ and Bob is given an input $y\in Y$, they need to output a value $z \in Z$ such that $(x,y,z)\in S$.
\end{defn}
\paragraph{Deterministic protocol}
\begin{defn}
	A deterministic protocol  $\Pi: {X} \times {Y} \rightarrow Z$ for a communication problem ${S} \subseteq \left({X} \times {Y}\right)\times Z$ is  a rooted binary tree with  following structure:
	\begin{itemize}
		\item  Every node $v$ in the tree  belongs to Alice or Bob and is associated with a rectangle $X_v \times Y_v \subseteq X\times Y$. Particularly, the root of protocol tree is associated with the  rectangle ${X} \times {Y}$.
		\item  Every internal node $v$ has two outgoing edges labeled with $0$ and $1$ respectively. These two edges labeled with $0$ and $1$  lead to $v$'s  two children  $v_0,v_1$ respectively.
		\item  Recall  $v$ is associated with a rectangle $X_v \times Y_v$, if $v$ is owned by Alice, then $v_0$ is associated with $X_{v_0} \times Y_v$, $v_1$ is associated with $X_{v_1} \times Y_v$ where $X_{v_0} \cap X_{v_1} = \emptyset$ and $X_{v_0} \cup X_{v_1} =X_v$; if $v$ is owned by Bob, then $v_0$ is associated with $X_v \times Y_{v_0}$, $v_1$ is associated with $X_v \times Y_{v_1}$ where $Y_{v_0} \cap Y_{v_1} = \emptyset$ and $Y_{v_0} \cup Y_{v_1} =Y_v$.
		\item  Every leaf node $\ell$ is  associated with a value $z\in Z$ as the output of the protocol. And for every leaf $\ell$, we have 
		$X_\ell \times  Y_\ell \times \{z\} \subseteq S$.
	\end{itemize}
\end{defn}

\begin{defn}Given a protocol tree $\Pi$ and a node $v$ in the tree,  the transcript of node $v$ is the string obtained by concatenating the labels of the edges in the path from the root to the node $v$. 
\end{defn}

\begin{defn}Given a protocol tree $\Pi$,  its depth $\mathsf{D}(\Pi)$ is the length of the longest path from the root to a leaf in the tree. Given a communication problem $S\subseteq (X\times Y) \times Z$, the  (deterministic) communication complexity $\mathsf{CC}(S)$ of communication problem $S$ is the minimum $\mathsf{D}(\Pi)$ over all protocol $\Pi$ for the problem $S$. 
\end{defn}

\paragraph{Non-Deterministic protocol}
\begin{defn}[Non-deterministic communication protocol \cite{DBLP:books/daglib/0011756,DBLP:conf/coco/MihajlinS21}]
	Given a function $f : X \times Y \rightarrow \{0,1\}$, we say  it has non-deterministic communication protocol of complexity $d$ if there are two functions $A : X \times \{0,1\}^d \rightarrow \{0,1\}$ and $B: Y \times \{0,1\}^d \rightarrow \{0,1\}$ such that
	\begin{itemize}
		\item $\forall (x, y) \in f^{-1}(1)\ \exists w \in \{0,1\}^d : A(x,w)=B(y,w)= 1$,
		\item  $\forall (x, y) \in f^{-1}(0)\ \forall w\in \{0,1\}^d : A(x,w)\neq 1 \text{ or } B(y,w)\neq 1$.
	\end{itemize}
	The  non-deterministic communication complexity of $f$ , denoted by $\mathsf{NCC}(f)$, is the minimal complexity over all non-deterministic communication protocols for $f$ . 
\end{defn}

\begin{defn}[Privately non-deterministic communication protocol \cite{DBLP:books/daglib/0011756,DBLP:conf/coco/MihajlinS21}]Given  a function $f : X \times Y \rightarrow \{0,1\}$,
we say it has privately non-deterministic communication protocol of complexity $d$ if there is a function $\hat{f} : (X \times \{0,1\}^\ast) \times  (Y \times \{0,1\}^\ast) \rightarrow \{0,1\}$ such that
\begin{itemize}
	\item $\forall (x, y) \in f^{-1}(1)\ \exists w_x, w_y\in \{0,1\}^\ast : \hat{f} ((x, w_x), (y, w_y)) = 1$,
	\item  $\forall (x, y) \in f^{-1}(0)\ \forall w_x, w_y\in \{0,1\}^\ast : \hat{f} ((x, w_x), (y, w_y)) = 0$,
\end{itemize}
and (deterministic) communication complexity of $\hat{f}$ is at most $d$.
The privately non-deterministic communication complexity of $f$, denoted by $\mathsf{NCC^\prime}(f)$, is the minimal complexity over all privately non-deterministic communication protocols for $f$ . 
\end{defn}

\begin{thm}[\cite{DBLP:conf/coco/MihajlinS21}]For any function  $f : X \times Y \rightarrow \{0,1\}$, we have
	$$\mathsf{NCC}(f) + 2 \geq  \mathsf{NCC^\prime}(f ) \geq \mathsf{NCC}(f ).$$
\end{thm}
\paragraph{Non-Deterministic complexity of non-equality problem}
\begin{defn}[The non-equality problem] Given a non-empty finite set $S$, the non-equality on $S$ is the function $\textnormal{NEQ}_S: S\times S \rightarrow \{0,1\}$ defined as follows: $\textnormal{NEQ}_S(x,y)=1$ if and only if $x\neq y$.
\end{defn}

\begin{fact}[\cite{DBLP:conf/coco/MihajlinS21}]\label{2.9}Given any non-empty finite set $S$, $\mathsf{NCC^\prime}(\textnormal{NEQ}_S) \geq \log \log |S|$.
\end{fact}

\subsection{Karchmer-Wigderson relations and their compositions}\label{s2.2}
We start by defining the universal relation and other involved Karchmer-Wigderson relations, then we define compositions of these relations.
\begin{defn}[Universal relation $\text{U}_n$]The Universal  relation $\text{U}_n$ is the following communication problem: Alice and Bob get inputs $x,y\in \{0,1\}^n$ respectively. Their task is to find a coordinate $i\in [n]$ such that $x_i \neq y_i$. If $x=y$, they can also output $\bot$.
\end{defn}

\begin{defn}[KW relation  over rectangle]Given two disjoint sets ${X},{Y} \subseteq \{0,1\}^n$, the KW relation over rectangle ${X}\times{Y}$, denoted by $\tKW_{{X}\times{Y}}$ is defined by
	\[\tKW_{{X}\times{Y}} \defeq \{(x,y,i) \mid x\in X, y\in Y, x_i \neq y_i\}.\]
\end{defn}

\begin{defn}[KW relation for functions]	Given a non-constant function $f:\{0,1\}^n \rightarrow \{0,1\}$, its KW relation $\tKW_{f}$ is defined by $\tKW_{f} \defeq  \tKW_{f^{-1}(1)\times f^{-1}(0)}$.
\end{defn}

\begin{defn}[The multiplexor relation $\text{MUX}_n$]In  KW relation $ \textnormal{MUX}_n$, Alice gets a function $f:\{0,1\}^{n} \rightarrow \{0,1\}$ and a Boolean string  $x\in \{0,1\}^{n}$,  Bob gets a function $g :\{0,1\}^{n} \rightarrow \{0,1\}$ and a Boolean string $y\in \{0,1\}^{n}$,  their goal is to find an entry $i$ such that $x_{i} \neq y_{i}$. If $f \neq g$ or $f(x)=g(y)$, they can also output $\bot$.
\end{defn}
\begin{rem}Here we want to point out  in the original version of multiplexor, the inputs to the players are promised to satisfy $f=g$ and $f(x)\neq g(x)$. Here	we use the rejectable version of multiplexor, that is when the promise is false, the players are allowed to reject and output $\bot$. The difference of the complexities of two versions is only two bits,  for example, Alice can send the $i$-th bit to  Bob, Bob replies with one bit that whether the answer $i$ is correct, if not, they output $\bot$. Thus, we ignore such difference and in the rest  of the paper, for problems similar to the multiplexor problem, we all present their rejectable versions.
\end{rem}

\begin{defn}[Composition of two Boolean functions and its KW relation]	Let $ f:\{0,1\}^{m} \rightarrow\{0,1\}$ and $g:\{0,1\}^{n} \rightarrow$ \{0,1\} be non-constant functions. The (block) composition of $f$ and $g$, denoted by $f \diamond g: \{0,1\}^ {m\times n}\rightarrow \{0,1\}$ , is defined as follows:
	\[f \diamond g(X) =f(g(X_1),\ldots,g(X_m)).\]
 In  KW relation $\tKW_{f \diamond g}$, Alice and Bob get $X \in(f \diamond g)^{-1}(1)$ and $Y \in(f \diamond g)^{-1}(0)$ viewed as $m \times n$ Boolean matrices, and their goal is to find an entry $(i, j)$ such that $X_{i, j} \neq Y_{i, j}.$ We also denote $\tKW_{f \diamond g}$ by $\tKW_{f} \diamond \tKW_{g}.$
\end{defn}
\begin{defn}[Composition of a universal relation and  a Boolean function]Let $\tU_m$ be the universal relation and  $f:\{0,1\}^{n}\rightarrow \{0,1\}$  be a function. Given any Boolean matrix $X\in \{0,1\}^{m \times n}$, define $f(X)=(f(X_1),\ldots,f(X_m))$. In KW relation $\tU_m \diamond \tKW_f$, Alice gets a Boolean matrix $X\in \{0,1\}^{m \times n}$, Bob gets a Boolean matrix $Y\in \{0,1\}^{m \times n}$, their goal is to find an entry $(i, j)$ such that $X_{i, j} \neq Y_{i, j}$. If $f(X)=f(Y)$, they can also output $\bot$.
\end{defn}

\begin{defn}[Composition of a universal relation and  a multiplexor]Let $\mathcal{F}$ be a set of functions $f : \{0,1\}^n \rightarrow \{0,1\}$.
	In  KW relation $\tU_m \diamond \textnormal{MUX}_\mathcal{F}$, Alice gets a function $f\in \mathcal{F}$ and a Boolean matrix $X\in \{0,1\}^{m \times n}$, Bob gets a function $g \in \mathcal{F}$ and a Boolean matrix $Y\in \{0,1\}^{m \times n}$,  their goal is to find an entry $(i, j)$ such that $X_{i, j} \neq Y_{i, j}$. If $f \neq g$ or $f(X)=g (Y)$, they can also output $\bot$.
\end{defn}

Now we make a detour to show how our idea emerges from the xor composition and finally avoids it. We start with the notion of generalized KW relations. Ivan Mihajlin and Alexander Smal \cite{DBLP:conf/isaac/IgnatievMS22} considered a more general form of KW relation including the case of non Boolean functions.
\begin{defn}[The generalized KW relation]Given a function $f:\{0,1\}^n \rightarrow \{0,1\}^k$, its KW relation $\tKW_{f}$ is defined by $\tKW_{f} \defeq  \{(x, y, i) \mid x, y \in \{0,1\}^n,f(x)\neq f(y),x_i \neq y_i\}$. The generalized KW relation $\tKW_{f}$  is the following communication problem: Alice and Bob get inputs $x,y\in \{0,1\}^n$ respectively. Their task is to find a coordinate $i\in [n]$ such that $x_i \neq y_i$ with the promise $f(x)\neq f(y)$.
\end{defn}

We will focus on a special form of generalized KW relation.
\begin{defn}[Function bundle]An $(m,n)$ function bundle  $F=(F_1,\ldots,F_m)$ is a tuple of $m$ functions $ F_1,\ldots,F_m:\{0,1\}^{n} \rightarrow \{0,1\}$ . We also treat $F$ as a function of $\{0,1\}^{m\times n} \rightarrow\{0,1\}^m$ defined as follows:
	$F(X)=(F_1(X_1),\ldots,F_m(X_m)).$
\end{defn}
Let $F$ be an $(m,n)$ function bundle, the generalized KW relation $\tKW_{F}$  is the following communication problem: Alice and Bob get two Boolean matrices $X,Y\in \{0,1\}^{m\times n}$ respectively. Their task is to find an entry $(i,j)$ such that $X_{i, j} \neq Y_{i, j}$ with the promise $F(X)\neq F(Y)$. The merit of studying this special form of generalized KW relation is illustrated in following fact which is implicit in \cite{DBLP:conf/coco/MihajlinS21}.

\begin{fact}\label{2.20}Given an  $(m,n)$ function bundle  $F$, define a function $h: \{ 0,1\}^{\log m+n} \rightarrow \{0,1\}$ such that
	$h(i,x) =F_i(x)$ where $i\in\{ 0,1\}^{\log m},x\in\{ 0,1\}^{n}$. Then, $\mathsf{CC}(\tU_m\diamond \tKW_h) \geq \mathsf{CC}(\tKW_F)$.
\end{fact}
We also can define a multiplexor of function bundles with restricted inputs.
\begin{defn}Let $\mathcal{F}$ be a set of function bundles and $\mathcal{X} \subseteq  \{0,1\}^{m\times n}$ . In a communication problem $\text{MUX}_\mathcal{F,X}$,  Alice gets  a function bundle $F \in \mathcal{F}$ and an  $X \in \mathcal{X}$, Bob gets  a function bundle $G \in \mathcal{F}$ and a $Y \in  \mathcal{X}$. Their goal is to find $(i,j)$ such that $X_{i,j} \neq Y_{i,j}$. If $F\neq G$ or $F(X) = G(Y)$, they can output $\bot$.
\end{defn}
Ivan Mihajlin and Alexander Smal introduced a so called xor composition  which is crucial for their results. They defined the xor composition of a universal relation and a multiplexor as follows.
\begin{defn}[\cite{DBLP:conf/coco/MihajlinS21}]In a communication problem $\text{U}_n \boxplus \text{MUX}_{n}^{\prime}$,  Alice is given a permutation function $F \in \{0,1\}^{n} \rightarrow \{0,1\}^{n}$ and two  strings $a,x\in  \{0,1\}^{n}$, Bob is given  a permutation function $G \in \{0,1\}^{n} \rightarrow \{0,1\}^{n}$ and two  strings  $b,y\in \{0,1\}^{n}$. Let $\circ$ be concatenation of strings and $\oplus$ be bit-wise xor. Their goal is to find
	$i\in [2n] $ such that $(a\circ x)_{i} \neq (b\circ y)_{i} $. If $F\neq G$ or $a\oplus F(x) = b \oplus G(y) $, they can output $\bot$.
\end{defn}

Ivan Mihajlin and Alexander Smal \cite{DBLP:conf/coco/MihajlinS21} proved $\mathsf{CC^{phd}}(\text{U}_n \boxplus \text{MUX}_{n}^{\prime}) \geq 1.5n-o(n)$ where $\mathsf{CC^{phd}}$ denotes the communication complexity in partially half-duplex model. Let's see that the above  xor composition $\text{U}_n \boxplus \text{MUX}_{n}^{\prime}$ can be viewed as a multiplexor of function bundles which take a restricted form of inputs.
\begin{fact}Let $\mathcal{P}$ be the set of all permutation functions over $n$ bit strings. Let $\mathcal{F}$ be a set of $(n,n+1)$ function bundles such that every $F=(F_1,\ldots,F_n) \in \mathcal{F}$ is generated from a permutation $G \in \mathcal{P}$.
Given a  permutation $G \in \mathcal{P}$, for every $i\in [n]$, define $F_i:\{0,1\}^{n+1}\rightarrow \{0,1\}$ to be $F_i(x\circ z)=G(x)_i \oplus z$ where $x\in \{0,1\}^n,z \in \{0,1\}$. Let $\mathcal{X}=\{X \in \{0,1\}^{n\times (n+1)}\mid x,a \in \{0,1\}^n, \forall i,X_i =x\circ a_i\}$. Then the communication problem $\text{MUX}_\mathcal{F,X}$ is essentially the same as the communication problem $\text{U}_n \boxplus \text{MUX}_{n}^{\prime}$.
\end{fact}
Our idea originates in trying to improve the xor composition theorem of  Ivan Mihajlin and Alexander Smal, soon we find out it is in fact a special form of multiplexor of function bundles. Then we  prove an almost tight lower bound of $\tKW_{F}$ for most function bundles. But by Fact \ref{2.20},  this only implies a lower bound of $\tU_m \diamond \tKW_h$ for many functions rather than for most functions. Finally, we manage to prove  the almost tight lower bound of $\tU_m \diamond \tKW_f$ for most functions  with a refined restriction technique.

\subsection{Half-duplex communication complexity}\label{s2.3}
To handle communication problems like the multiplexor problem, Hoover, Impagliazzo, Mihajlin and Smal \cite{DBLP:conf/isaac/HooverIMS18} proposed a generalization of the classical communication model, the half-duplex model. Unlike Yao's classical model of communication \cite{DBLP:conf/stoc/Yao79}, in  each round of the half-duplex model, the players can synchronize their clocks and perform actions simultaneously. At the beginning of one clock cycle,   each player takes one of three actions: send $0$, send $1$, or receive. If  the action of receive is taken, the player  listens to the communication channel and receives one bit at the end of the clock cycle.  Thus at the end of every round, each player will eventually perform  one of four actions: receive $0$ ($\texttt{r(0)}$ for short), receive $1$ ($\texttt{r(1)}$ for short), send $0$ ($\texttt{s(0)}$ for short), and send $1$ ($\texttt{s(1)}$ for short). Let $\texttt{Action = \{r0,r1,s0,s1\}}$ be the set of all such actions. According to the actions taken by the players, intuitively, there are three different kinds of rounds: \emph{classical}, \emph{wasted} and \emph{silent}.
\begin{itemize}
	\item In a \emph{classical} round,  one player sends some bit and the other one receives  such bit as the case in the classical model of communication.
	\item In a \emph{wasted} round,  both players send bits  but since no one is listening,  such bits never get received thus wasted.
	\item In a \emph{silent} round,  both players receive.  Since no one is actually speaking, the channel is silent in this round.
\end{itemize}
Here the tricky thing is about the silent round, since at the end of a silent round, both players eventually receive certain bits those neither of players send. There are different ways \cite{DBLP:conf/isaac/HooverIMS18} to determine those bits received in a silent round, in this paper, we focus on the model of half-duplex communication with adversary  where in silent round both players receive certain bits which are chosen by an adversary. Formally, we have following definition.
\begin{defn}[Half-duplex protocol with adversary]
	A deterministic half-duplex protocol with adversary  $\Pi: {X} \times {Y} \rightarrow Z$ for a communication problem ${S} \subseteq \left({X} \times {Y}\right)\times Z$ 
	is a  pair of full $4$-ary trees $(\Pi_A,\Pi_B)$ with the same depth $d$ owned by Alice and Bob respectively. And two trees $\Pi_A,\Pi_B$ are with the following structure:
\begin{itemize}
	\item  Every node $v$ in the tree $\Pi_A$ (respectively $\Pi_B$) is associated with a subset $X_v \subseteq X$ (respectively $Y_v \subseteq Y$). Particularly, if the node $v$ is the root of $\Pi_A$ (respectively $\Pi_B$), it is associated with $X$ (respectively $Y$). Each node $v$ represents certain state in the tree, a  state pair $(u,v)$ from two trees $\Pi_A,\Pi_B$  represents certain state  in the protocol $\Pi$, thus we will also treat $u,v$ as states. The subset $X_v$ (respectively $Y_v$) is the set of inputs to Alice (respectively Bob) that can reach the node $v$ from the root of   tree $\Pi_A$ (respectively $\Pi_B$).
	\item  Every internal node $v$ has $4$ outgoing edges. Each edge is  labeled with one action $\texttt{ac}$ from $\texttt{\{r(0),r(1),s(0),s(1)\}}$ respectively. Each edge labeled with the action  $\texttt{ac}$  leads to $v$'s  child  $v_{\texttt{ac}}$.
	\item  Recall that each node $v$  of tree $\Pi_A$ is associated with a subset  $X_v \subseteq X$, then $X_v$ is partitioned into three disjoint subsets $X_{v:\texttt{r}},X_{v:\texttt{s(0)}},X_{v:\texttt{s(1)}}$. Similarity, every node $v$ in the tree $\Pi_B$ is associated with a subset  $Y_v \subseteq Y$  and  $Y_v$  is partitioned into three disjoint subsets $Y_{v:\texttt{r}},Y_{v:\texttt{s(0)}},Y_{v:\texttt{s(1)}}$. The partition of inputs indicates what action the player takes at the beginning of each round. 
	\item  Every leaf node $\ell$ is  associated with a value $z_\ell\in Z$ as the output of the protocol. 
\end{itemize}
Now let's see how the protocol find out the answer to  any input $(x,y)\in X\times Y$.  The protocol maintains a pair of states $(u,v)$ where $u,v$ are nodes at the same depth in the trees $\Pi_A,\Pi_B$. Alice knows the state $u$ and the input $x$, meanwhile Bob   holds the state $v$ and the input $y$. Initially, $(u,v)$ are the two roots of  trees $\Pi_A,\Pi_B$. When  $u,v$ are not leaves in  the trees $\Pi_A,\Pi_B$, the protocol takes some  action from following cases and updates the pair of states until $u,v$ are leaves.
\begin{itemize}
	\item  If $x\in X_{u:\texttt{s(b)}}$ for some   $\texttt{b} \in \{0,1\}$ and  $y\in Y_{v:\texttt{r}}$, Alice sends a bit $\texttt{b}$ and Bob receives such bit $\texttt{b}$. The protocol updates the state pair $(u,v)$ to new state pair $(u_{\texttt{s(b)}},v_{\texttt{r(b)}})$. This is a \emph{classical} round.
	\item Similarly, if $x\in X_{u:\texttt{r}}$ and  $y\in Y_{v:\texttt{s(b)}}$ for some   $\texttt{b} \in \{0,1\}$, Bob sends a bit $\texttt{b}$ and Alice receives such bit $\texttt{b}$. The protocol updates the state pair $(u,v)$ to new state pair $(u_{\texttt{r(b)}},v_{\texttt{s(b)}})$. This is  also a \emph{classical} round.
	\item If $x\in X_{u:\texttt{s(b)}}$  for some   $\texttt{b} \in \{0,1\}$  and  $y\in Y_{v:\texttt{s(d)}}$ for some   $\texttt{d} \in \{0,1\}$, Alice sends a bit $\texttt{b}$ and Bob sends a bit $\texttt{d}$. The protocol updates the state pair $(u,v)$ to new state pair $(u_{\texttt{s(b)}},v_{\texttt{s(d)}})$. This is a \emph{wasted} round.
	\item If $x\in X_{u:\texttt{r}}$ and  $y\in Y_{v:\texttt{r}}$, the adversary chooses two bits $\texttt{b,d} \in \{0,1\}$,  Alice receives bit $\texttt{b}$ and Bob receives  bit $\texttt{d}$. The protocol updates the state pair $(u,v)$ to new state pair $(u_{\texttt{r(b)}},v_{\texttt{r(d)}})$. This is a \emph{silent} round.
\end{itemize}
When the protocol finally reaches a pair of states $(u,v)$ where $u,v$ are leaves of trees $\Pi_A,\Pi_B$ respectively, the protocol outputs the result $(z_u,z_v)$. We say the protocol $\Pi$ (correctly) solves the problem $S\subseteq X\times Y \times Z$, if for every input $(x,y)$, the protocol $\Pi$ reaches  some pair of states $(u,v)$ where $u,v$ are leaves  such that $z_u=z_v=z$ and $(x,y,z)\in S$ no matter what bits are chosen by the adversary in any silent round. The complexity $\mathsf{CC^{hd}}(\Pi)$ of the protocol $\Pi$  is the depth $d$ of two trees $\Pi_A, \Pi_B$, recall that we require two trees $\Pi_A, \Pi_B$ are of the same depth $d$. The  deterministic communication complexity of $S$ in half-duplex model with adversary, denoted by $\mathsf{CC^{hd}}(S)$, is the minimal complexity over all  deterministic half-duplex protocol with adversary  for $S$. 
\end{defn}
Now we introduce some useful notations and facts. The first notion is the legal action pair. Recall that in every round, eventually Alice and Bob take some $\texttt{ac}_A,\texttt{ac}_B \in \texttt{Action}$ respectively, those two actions $\texttt{ac}_A,\texttt{ac}_B$ form an action pair $(\texttt{ac}_A,\texttt{ac}_B)$. But not every action pair  from $\texttt{Action} \times \texttt{Action}$ is legal, particularly  in the classical round, the bit sent must be the same as the bit received, thus action pairs such as  $$(\texttt{s(1),r(0)}),(\texttt{s(0),r(1)}),(\texttt{r(1),s(0)}),(\texttt{r(0),s(1)})$$ are all \emph{illegal}. Now let $\sigma$ be a sequence of legal action pairs, let 
$\sigma_A$ (respectively $\sigma_B$) be a sequence of actions taken by Alice (respectively Bob), then $\sigma$ determines a unique legal state pair $(u,v)$ where $u,v$ are determined by $\sigma_A,\sigma_B$ in two tree $\Pi_A,\Pi_B$ respectively. Indeed, given a sequence $\sigma_A$ of actions taken by Alice, $\sigma_A$ defines the unique path from the root to $u$, the case for $v$ is similar. Now we try to define the  transcript in  the model of half-duplex communication with adversary and make it compatible to the transcript in the classical model of communication.  Given a sequence $\sigma_A$ of actions taken by Alice, let the $\pi(\sigma_A)$ be the ordered  bits involved in the actions, we say  $\pi(\sigma_A)$ is Alice's transcript. Similarity, let the $\pi(\sigma_B)$ be the ordered bits involved in the actions taken by Bob, we say  $\pi(\sigma_B)$ is Bob's transcript. But Alice's transcript is not always consistent with the one  of Bob, thus in general, we can not have transcript for the entire protocol. Nevertheless, if all action pairs in a sequence are classical, we can have a consistent transcript for both players. Formally,  We have following definitions.
\begin{defn}Let  $(\texttt{ac}_A,\texttt{ac}_B)$ be an action pair taken in a classical round, we say it is a classical action pair. If an action pair sequence $\sigma$ contains only classical action pairs, we say the sequence $\sigma$ is classical. Let  $(u,v)$ be the state pair determined by a  classical sequence $\sigma$ of action pairs,  we say  $(u,v)$ is a classical state pair. Let  $\sigma$ be a classical sequence of action pairs and $(u,v)$ be the state pair determined by $\sigma$, let  $\pi \in \{0,1\}^\ast$ be the ordered  bits  involved in   sequence  $\sigma$,  we say $\pi$ is a  protocol's transcript. Furthermore, we say both $\sigma$ and  $(u,v)$ are consistent with  protocol's transcript $\pi$.  Note that due to  different choices of the sender,   there may be several classical state pairs at given depth such that all of them  are consistent with one same protocol's transcript. For simplicity, if we say $\pi$ is a  transcript, we mean it's a protocol's transcript rather than some player's transcript.
\end{defn}


In  classical model of communication, one important property is the rectangle property. That is there is a rectangle associated with each node $v$ in the protocol tree. But this is not true in  half-duplex model with adversary, due to the interference of  the adversary. In general, it is not true that for every state pair $(u,v)$ the inputs which reach $(u,v)$  form a rectangle. Nevertheless, if we concern the classical state pair, the rectangle property is true. 

\begin{defn}Given a input $(x,y)$, if the protocol reaches  a state pair  $(u,v)$ along some sequence $\sigma$ of action pairs, we say the input $(x,y)$ is consistent with the state pair  $(u,v)$. More over if the state pair  $(u,v)$ is consistent with a transcript $\pi$,  we say input $(x,y)$ is also consistent with the transcript $\pi$. 
\end{defn}

\begin{rem}Note that due to the adversary, one input $(x,y)$ may be consistent with several distinct state pairs at  given depth, but one input $(x,y)$ can only be consistent with at most one classical state pair at given depth and one protocol's transcript of given length, since the adversary can not interfere any classical round. 
\end{rem}


Now we show the rectangle property is true for every classical state pair. 
\begin{fact}\label{2.28}Given two input pairs $(x,y),(x^\prime,y^\prime)$, if both $(x,y),(x^\prime,y^\prime)$ are consistent with some classical state pair $(u,v)$, then input pairs $(x,y^\prime),(x^\prime,y)$ are also consistent with the classical state pair $(u,v)$.
\end{fact}
\begin{proof}We prove this fact by induction on the depth  of the state pair. Initially, $(u,v)$ are roots of   two trees $\Pi_A, \Pi_B$, $X$(respectively $Y$) is associated with root $u$(respectively $v$), if both $(x,y),(x^\prime,y^\prime)$ are consistent with the classical state pair $(u,v)$, $x,x^\prime \in X$ and $y,y^\prime \in Y$, thus $(x,y^\prime),(x^\prime,y)$ are also consistent with the classical state pair $(u,v)$.  Now assume  there are two input pairs $(x,y),(x^\prime,y^\prime)$ which are consistent with some classical state pair $(u^\prime,v^\prime)$, and let $u, v$ be parent nodes of $u^\prime,v^\prime$ respectively. W.l.o.g., assume the classical state pair $(u, v)$ transits to $(u^\prime,v^\prime)$ via action pair $(\texttt{s(0),r(0)})$. Since $(x,y),(x^\prime,y^\prime)$  are consistent with  $(u^\prime,v^\prime)$, they must be also consistent with  $(u,v)$ in the first place, by induction hypothesis, $(x,y^\prime),(x^\prime,y)$  are consistent with  $(u,v)$, we will show $(x,y^\prime),(x^\prime,y)$  are also consistent with  $(u^\prime,v^\prime)$. Now since $(x,y),(x^\prime,y^\prime)$  are consistent with  $(u^\prime,v^\prime)$, $x, x^\prime \in X_{u:\texttt{s(0)}}$ and   $y, y^\prime \in Y_{v:\texttt{r}}$. Therefore, given input pairs $(x,y^\prime),(x^\prime,y)$ at state pair $(u,v)$, after Alice  and Bob take actions $\texttt{s(0),r(0)}$ respectively, the protocol also enters state $(u^\prime,v^\prime)$, thus $(x,y^\prime),(x^\prime,y)$ are also consistent with $(u^\prime,v^\prime)$ as required.
\end{proof}

\paragraph{Partially half-duplex communication}
When we handle a communication problem similar to the multiplexor, we consider a more restricted model of half-duplex  communication with adversary which is called the \emph{partially half-duplex} communication model. In such model, each player's input contains two parts: Alice gets $(f,x)$ and Bob gets $(g,y)$. They can use a half-duplex protocol for their task but not with its full power, when $f=g$, the protocol is only allowed to perform  classical rounds. We use $\mathsf{CC^{phd}}$ to denote the communication complexity of a problem in partially half-duplex model with adversary.

\begin{fact}\label{2.29}Let $\Pi$ be a partially half-duplex protocol for some communication problem and the depth of $\Pi$ is at least $d$. Let $\mathcal{D}$ be a set of inputs to the protocol and every input to the protocol in $\mathcal{D}$ is of form $((f,x),(f,x^\prime))$, then there is a transcript $\tau \in  \{0,1\}^d$ and a subset $\mathcal{D}^\prime \subseteq \mathcal{D}$ such that $|\mathcal{D}^\prime|  \geq |\mathcal{D}|/2^d$ and  every input in $\mathcal{D}^\prime$ is consistent with the transcript $\tau$.
\end{fact}
\begin{proof}Since the depth of $\Pi$ is at least $d$, there must be  transcripts of length $d$. Given any fixed input $((f,x),(f,x^\prime))$ in $\mathcal{D}$, since  $\Pi$ is partially half-duplex, $((f,x),(f,x^\prime))$ must be consistent with some transcript $\tau$  of length $d$. Moreover, there are at most $2^d$ such transcripts, there must be one $\tau$ and a subset $\mathcal{D}^\prime \subseteq \mathcal{D}$ of size at least $|\mathcal{D}|/2^d$ such that every input in $\mathcal{D}^\prime$ is consistent with $\tau$.
\end{proof}

\section{A Composition Theorem of a Universal Relation and a Multiplexor }\label{s3}
In this section, we prove the lower bound for $\textnormal{U}_m \diamond \textnormal{MUX}_\mathcal{F}$ in  the model of partially half-duplex communication with adversary. At first, let's see the overall strategy of the proof. When $n\geq m$, we can use a two-stage argument to show that after the protocol has spent approximate $m$ bits communication, it still needs another approximate $n$ bits to completely solve the problem. After spent approximate $m$ bits, we can extract a set of inputs from the residual problem and use it to solve the non-equality problem of size approximate $2^{2^n}$ non-deterministically thus the protocol needs another approximate $n$ bits. 

But when $n$ is much smaller than $m$,  there is some subtle issue about this argument. In order to apply the two stage argument we must be able to show the protocol needs to spend about $m$ bits in the first stage, but now we  are only able to show that the protocol needs to spend about $n$ bits in the first stage. Nevertheless, we can repeatedly use the two stage argument  to boost the complexity of the protocol up until it's done. In general, we can show a boosting theorem that  is after  the protocol has spent  $s\leq m-o(m)$ bits in  first stage, it still requires another approximate $n$ bits to compete the task. We can repeatedly use the boosting theorem to add approximate  $n$ to $s$ until $s$ is about $m$, then we add a final approximate $n$ to $s$ and obtain the final complexity which is about $m+n$. 

More formally, the boosting theorem depends on two following lemmas: a boosting  lemma and an extraction lemma.  Let $\epsilon,c,t$ be parameters which depend on $m,n$. Assume a protocol $\Pi$ has spent $s\leq  m-t-1$ bits, let $\mathcal{S}$ be a subset of $\mathcal{F} \times \mathcal{X}$, the residual protocol has to solve every input of form $((f,X),(f,X))$ where $(f,X)\in \mathcal{S}$. The extraction lemma allows us to extract a set of function $\mathcal{H}$ of size  at least $2^{2^{(1-\epsilon)n}}$ such that  for all distinct $f, g \in \mathcal{H}$, there exists  an $ X:(f,  X), (g,  X)\in \mathcal{S}$, and $f(X) \neq  g(X)$. Then the boosting lemma can use the set  $\mathcal{H}$ and the protocol $\Pi$ to solve  $\text{NEQ}_\mathcal{H}$ with a privately non-deterministic communication protocol, thus the protocol $\Pi$ will need another (approximate) $\log\log |\mathcal{H}|$ bits communication. To proceed, we need following definition which treats any subset  $\mathcal{Z}\subseteq \mathcal{F}\times  \mathcal{X}$  as a bipartite graph.
\begin{defn} Let $\mathcal{F}$ be a set of functions $f : \{0,1\}^n \rightarrow \{0,1\}$ and $\mathcal{X}$ be the set $\{0,1\}^{m\times n}$. Given a set $\mathcal{Z}\subseteq \mathcal{F}\times  \mathcal{X}$, we define the domain graph $\Gamma_\mathcal{Z}$ to be a bipartite graph
	$(\mathcal{U}_\mathcal{Z}, \mathcal{V}_\mathcal{Z},\mathcal{E}_\mathcal{Z})$, such that $\mathcal{U}_\mathcal{Z} =\{f \mid (f,X)\in \mathcal{Z}\}$, $\mathcal{V}_\mathcal{Z} =\{X \mid (f,X)\in \mathcal{Z}\}$, and $(f, X)\in \mathcal{E}_\mathcal{Z} \iff (f, X)\in \mathcal{Z}$. Furthermore,  for every $f\in \mathcal{U}_\mathcal{Z}$,  denote $\{X\mid X\in \mathcal{X}, (f,X) \in \mathcal{Z}\}$ by $\mathcal{X}_{\mathcal{Z},f}$.
\end{defn}





Now we prove the boosting lemma, its idea is similar to that in  \cite{DBLP:conf/coco/MihajlinS21}, we adapt their idea to our case and present a more detailed proof.
\begin{lem}[The boosting lemma]\label{3.2}Let  $\mathcal{F}$ be a set of functions $f : \{0,1\}^n \rightarrow \{0,1\}$, $\mathcal{X}$ be the set $\{0,1\}^{m\times n}$ and  $\Pi$ be a partially half-duplex communication protocol solving $\textnormal{U}_m \diamond \textnormal{MUX}_\mathcal{F}$. Assume the protocol $\Pi$ has spent $s\leq m$ rounds communication, let  $\tau \in \{0,1\}^s$ be a partial transcript, there is a set $\mathcal{S} \subseteq \mathcal{F} \times \mathcal{X}$ such that every input from $\{((f,X),(f,X))\mid (f,X) \in \mathcal{S}\}$ is consistent with the transcript $\tau$. Let $\Gamma_\mathcal{S} = (\mathcal{U}_\mathcal{S}, \mathcal{V}_\mathcal{S},\mathcal{E}_\mathcal{S})$ be the domain graph of $\mathcal{S}$,  if there is  a set $\mathcal{H}\subseteq \mathcal{U}_\mathcal{S}$  such that for all distinct $f, g \in \mathcal{H}$, there exists an $ X:(f,  X), (g,  X)\in \mathcal{S}$, and $f(X) \neq  g(X)$, let $d$ be the depth of  $\Pi$, assume $d-s\leq n$, then  $d\geq  s+\log\log \mathcal{H}-\log m -\log n -6$.
\end{lem}
\begin{proof} Let $\mathcal{S} \subseteq \mathcal{F} \times \mathcal{X}$  be the set such that every input from $\{((f,X),(f,X))\mid (f,X) \in \mathcal{S}\}$ is consistent with the transcript $\tau \in \{0,1\}^s$, we will show how to use the protocol to solve  $\text{NEQ}_\mathcal{H}$ with a privately non-deterministic communication protocol.  Given  $f,g \in \mathcal{H}$, Alice and Bob check one of following three conditions to be true to make sure $f\neq g$. If $f=g$, all these conditions are false.
\begin{itemize}	
		\item The first condition is  there exist  $X \in \mathcal{X}_{\mathcal{S},f}, Y \in \mathcal{X}_{\mathcal{S},g}$ such that  $((f,X), (f,X)), ((g,Y), (g,Y))$ are consistent with two distinct state pairs at depth $d$ respectively, meanwhile both two distinct state pairs  are consistent with the transcript $\tau$. 
		\item The second  condition is there exist  $X \in \mathcal{X}_{\mathcal{S},f}, Y \in \mathcal{X}_{\mathcal{S},g}$ such that $f(X) \neq g(Y)$ and the residual protocol  performs at least one non-classical round to solve $((f,X),(g,Y))$.
		\item Finally, the third condition is there exist  $X \in \mathcal{X}_{\mathcal{S},f}, Y \in \mathcal{X}_{\mathcal{S},g}$ such that $f(X) \neq g(Y)$ and the residual protocol solves $((f,X),(g,Y))$ with  only classical rounds and outputs $\bot$.

	\end{itemize}
       Now we give a detailed description of  the  privately non-deterministic communication protocol to solve  $\text{NEQ}_\mathcal{H}$. When Alice gets a function $f\in \mathcal{H}$ and Bob gets a function $g\in \mathcal{H}$,  at first Alice guesses one condition out of the three and tells Bob with $2$ bits communication which condition they are going to verify, then they verify that condition as follows.
       \begin{itemize}
       	\item  For the first condition, Alice guesses an $X \in \mathcal{X}_{\mathcal{S},f}$ then Alice simulates the protocol $\Pi$ on input $((f,X),(f,X))$ and obtains  a sequence $\sigma$ of classical action pairs  which is consistent with $\tau$. Then Alice guesses an index $i\in [s]$; Bob guesses a $Y \in \mathcal{X}_{\mathcal{S},g}$,  then  simulates the protocol $\Pi$ on input $((g,Y),(g,Y))$ and obtains  a sequence $\sigma^\prime$ of classical action pairs  which is consistent with $\tau$. Alice sends $i$ to Bob and uses another $1$ bit to tell Bob who sends in $i$-th round of $\sigma$. If the $i$-th round of $\sigma$ is consistent with the $i$-th round of $\sigma^\prime$, Bob replies  Alice with $0$. Otherwise  Bob sends $1$ to Alice.  To check the first condition requires at most $\log m +2$ bits communication.
       	\item For the second condition,  Alice guesses an $X \in \mathcal{X}_{\mathcal{S},f}$, then simulates the protocol $\Pi$ on input $((f,X),(f,X))$ and obtains  a sequence $\sigma$ of classical action pairs  which is consistent with $\tau$, let $(u,v)$ be the state pair the protocol reaches.  Bob guesses a $Y \in \mathcal{X}_{\mathcal{S},g}$, then simulates the protocol $\Pi$ on input $((g,Y),(g,Y))$ and obtains  a sequence $\sigma^\prime$ of classical action pairs  which is consistent with $\tau$, let $(u^\prime,v^\prime)$ be the state pair the protocol reaches. Now Alice guesses a number $s^\prime \in [d-s]$, a string $\tau^\prime \in \{0,1\}^{s^\prime}$, a coordinate $i \in [m]$ and two bits $\mathtt{a \in \{receive,send\},b}\in\{0,1\}$, then sends all $s^\prime,\tau^\prime, i,\mathtt{a,b}$ to Bob, and they verify following to be true.
       	\begin{itemize}       	 
       		\item $\mathtt{b}=f(X)_{i}\neq g(Y)_{i}=1-\mathtt{b}$.
       		\item Alice simulates the protocol from node $u$ in the tree $\Pi_A$ according to the string $\tau^\prime$, that is each bit involved in each action  must be consistent with the corresponding bit in $\tau^\prime$. Similarly, Bob simulates the protocol from node $v^\prime$ in the tree $\Pi_B$ according to the string $\tau^\prime$.  After $s^\prime$ rounds, Alice and Bob   verify the actions they take in next round are the same as the bit $\mathtt{a}$ indicates: either both receive or both send.
       	\end{itemize}
       After all that, Alice and Bob use two bits communication to tell each other the results. The second condition requires at most $d-s+\log m +\log n +4 $ bits communication.

       	\item For the third condition, similarly, Alice guesses an $X \in \mathcal{X}_{\mathcal{S},f}$, then simulates the protocol $\Pi$ on input $((f,X),(f,X))$ and obtains  a sequence $\sigma$ of classical action pairs  which is consistent with $\tau$, let $(u,v)$ be the state pair the protocol reaches.  Bob guesses a $Y \in \mathcal{X}_{\mathcal{S},g}$, then simulates the protocol $\Pi$ on input $((g,Y),(g,Y))$ and obtains  a sequence $\sigma^\prime$ of classical action pairs  which is consistent with $\tau$, let $(u^\prime,v^\prime)$ be the state pair the protocol reaches. Now Alice guesses a string $\tau^\prime \in \{0,1\}^{d-s}$, a coordinate $i \in [m]$ and a bit $\mathtt{b}\in\{0,1\}$, then sends all $\tau^\prime,i,\mathtt{b}$ to Bob, and they verify following to be true.
       	\begin{itemize}       	 
       		\item $\mathtt{b}=f(X)_{i}\neq g(Y)_{i}=1-\mathtt{b}$.
       		\item Alice simulates the protocol from node $u$ in the tree $\Pi_A$ according to the string $\tau^\prime$ meanwhile Bob simulates the protocol from node $v^\prime$ in the tree $\Pi_B$ according to the string $\tau^\prime$.  After $s^\prime$ rounds, Alice and Bob verify they both reach leaves labeled with $\bot$.
       	\end{itemize}
       	After all that, Alice and Bob use two bits communication to tell each other the results. The third condition requires  $d-s+\log m+3$ bits communication.
       \end{itemize}
Now we show this privately non-deterministic  protocol is correct. 	Suppose that $f = g$. Then neither of three conditions could be true.  Since $f = g$ the protocol behaves as a classical one, any  transcript determines who sends in each round. Now the transcript $\tau$ is fixed already,  the sequence of action pairs is the same for every $((f,X),(f,X)),X \in \mathcal{X}_{\mathcal{S},f}$, thus the first condition is false. 
By the definition of partially half-duplex protocol and $f=g$, the second condition is also false. For every input $((f,X),(f,Y)),X,Y\in \mathcal{X}_{\mathcal{S},f},f(X)\neq f(Y)$, the protocol $\Pi$ should output $(i,j)$ such that  $X_{i,j}\neq Y_{i,j}$ rather than $\bot$, it means the third condition also fails.

Suppose that $f \neq g$. If the first or the second condition is true, then we have $f\neq g$ already.  If this is not the case, the third condition must be true. Now since the first condition is false, that is for  every $X \in \mathcal{X}_{\mathcal{S},f},Y \in \mathcal{X}_{\mathcal{S},g}$, the protocol takes the same sequence of classical action pairs  upon inputs $((f,X),(f,X)), ((g,Y),(g,Y))$ and $((f,X),(f,X)), ((g,Y),(g,Y))$ are consistent with the same classical state pair $(u,v)$. By  the rectangle property of classical state pair of Fact \ref{2.28}, for every $X \in \mathcal{X}_{\mathcal{S},f},Y \in \mathcal{X}_{\mathcal{S},g}$, $((f,X),(g,Y))$ is also consistent with $(u,v)$. Let $\mathcal{R}_{f,g}$ be the set $\{((f,X),(g,Y)) \mid X \in \mathcal{X}_{\mathcal{S},f},Y \in \mathcal{X}_{\mathcal{S},g}\}$, this means every input in $\mathcal{R}_{f,g}$ will be solved correctly by the residual protocol starting at   $(u,v)$. Let $\mathcal{R}^\prime _{f,g}$ be the set $\{((f,X),(g,Y)) \mid X \in \mathcal{X}_{\mathcal{S},f},Y \in \mathcal{X}_{\mathcal{S},g}, f(X)\neq g(Y)\}$ and since for every $f,g$ there exists an $X^\star$ such that $f(X^\star)\neq g(X^\star)$, $\mathcal{R}^\prime _{f,g}$ is not empty. When the second condition is also false,  it means the residual protocol solves every input from $\mathcal{R}^\prime _{f,g}$ correctly with only classical rounds. By the definition of $\textnormal{U}_m \diamond \textnormal{MUX}_{\mathcal{F}}$, to correctly solve $((f,X^\star),(g,X^\star))$ Alice and Bob must reach leaves labeled with $\bot$ as required.

 The total number of bits communicated in the privately non-deterministic  protocol is at most $d-s+\log m+\log n+6$.  By Fact \ref{2.9},  $d-s+\log m+\log n+6\geq \log\log |\mathcal{H}|$, thus $d\geq  s+\log\log  |\mathcal{H}|-\log m-\log n-6$.
\end{proof}
\begin{rem}Note that the string $\tau^\prime$ is necessary, Alice and Bob use the common string $\tau^\prime$ to make sure  in every classical round the bits in their actions are consistent.  Without the common  string, there may be illegal action pairs.
\end{rem}


\begin{lem}[The extraction lemma]\label{3.4}Let $m,n$ be integers  such that $m\geq 1,n> 2\log m+2$. Let $\epsilon \in ( \frac{\log m +2}{n},1-\frac{\log m}{n})$ be a parameter and $c,t$ be integers satisfying $c\geq \frac{2m+\log m}{\epsilon n-\log m -2},t\geq c+4$. Let $\mathcal{F}$ be a set of functions $f:
	\{0,1\}^n \rightarrow \{0,1\}$ such that $|\mathcal{F}|   \geq 2^{-2^{(1-\epsilon)n}} \cdot 2^{2^n}$. Let $\mathcal{X}$ be the set $\{0,1\}^{m\times n}$.
Let $\mathcal{S}\subseteq \mathcal{F}\times \mathcal{X}$ be a subset such that $|\mathcal{S}| \geq 2^{t-m}\cdot |\mathcal{F}|\cdot |\mathcal{X}|$, and let $\Gamma_\mathcal{S} = (\mathcal{U}_\mathcal{S}, \mathcal{V}_\mathcal{S},\mathcal{E}_\mathcal{S})$ be the domain graph of $\mathcal{S}$,  for every $f\in \mathcal{U}_\mathcal{S}$,   $|\mathcal{X}_{\mathcal{S},f}| \geq 2^{t-m}\cdot |\mathcal{X}|$.
Then there is   a set $\mathcal{H}\subseteq \mathcal{U}_\mathcal{S}$ of size at least $2^{2^{(1-\epsilon)n}}$  such that for all distinct $f, g \in \mathcal{H}$, there exists an $ X:(f,  X), (g,  X)\in \mathcal{S}$, and $f(X) \neq  g(X)$.
\end{lem}
\begin{proof}We extract the $\mathcal{H}$ from $\mathcal{U_S}$ by constructing a tree $T(\mathcal{S}^\prime)$ rooted with $\mathcal{S}^\prime \subseteq \mathcal{S}$ such that
	\begin{itemize}
	\item in the tree, each  node $z$  is associated with a subset $\mathcal{Z} \subseteq \mathcal{S}$ which viewed as a domain graph $\Gamma_\mathcal{Z} = (\mathcal{U}_\mathcal{Z}, \mathcal{V}_\mathcal{Z},\mathcal{E}_\mathcal{Z})$ and if $z$ is not a leaf, internal node $z$ is also labeled with an $X_\mathcal{Z}\in \mathcal{V_Z}.$ Sometimes, to emphasize they are associated with node $z$, we also denote them with subscript $z$ such as   $\mathcal{U}_z,\mathcal{V}_z,\mathcal{E}_z$ and $X_z$.
	\item For every two distinct leaves $\ell_1,\ell_2$,  we have $\mathcal{U}_{\ell_1}\cap \mathcal{U}_{\ell_2} =\emptyset$. Let node  $v$ be  the lowest common ancestor of these two leaves and $v$ is  labeled with
		$X$, then for all $f\in \mathcal{U}_{\ell_1}, g \in \mathcal{U}_{\ell_2}$, we have $(f,  X), (g,  X)\in \mathcal{S}$  and $f(X) \neq  g(X)$.
	\end{itemize}
After the  tree $T(\mathcal{S}^\prime)$ is constructed, the set $\mathcal{H}$ is obtained by taking exact one function from each leaf.  Given two distinct elements $f, g \in \mathcal{H}$ such that $f\in \mathcal{U}_{\ell_1}, g\in \mathcal{U}_{\ell_2}$, since $\mathcal{U}_{\ell_1}\cap \mathcal{U}_{\ell_2} =\emptyset$, $f \neq  g$. Moreover, let $X$ be the label of the least common ancestor of  leaves $\ell_1$ and $\ell_2$, we have  $(f,  X), (g,  X)\in \mathcal{S}$ and $ f(X) \neq  g(X)$ as required.

Before construction of  the tree, we need to introduce some helpful notations. A trace $\Psi$ is a subset of $\{0,1\}^n$. Particularly, we can view every $X \in \{0,1\}^{m\times n}$ as a trace, for convenience, when the context is clear, we abuse the notation and treat $X$ as  a trace set of all its distinct rows $\{x\mid \exists i, x =X_i\}$.  Let $z$ be a node at depth $d$ in the tree, from root to node $z$, its ancestors are $z_0,z_1,\ldots,z_{d-1}$. For every $i\in \{0,1,\ldots,d-1\}$, $z_i$ is labeled with  $X_{z_i}$, treat every $X_{z_i}$ as a trace, we define $\Psi(z) = \bigcup_{i=0}^{d-1} X_{z_i}$.

The purpose of   trace  is to record a set of inputs $\Psi(z)$ and all  functions in $\mathcal{U}_z$ take the same value given any input in  $\Psi(z)$. In another word, all functions in $\mathcal{U}_z$ are restricted to set $\{0,1\}^n \setminus \Psi(z)$. Therefore, the number of functions in $\mathcal{U}_z$ is up bounded by $2^{2^n-|\Psi(z)|}$. For our purpose, we need the size of $\mathcal{U}_z$ to be as small as possible thus the size of trace $\Psi(z)$ to be as large as possible. Given a trace $\Psi(z)$ for some node $z$, we want to choose  an $X$ for node $z$ such that $|\Psi(z)\cup X| -|\Psi(z)| \geq m-c$. The problem is that we can not choose any $X$ freely, to make any remaining $X$ is good for our purpose, we have to remove all bad $X$s in advance. Let 
$$\Phi(z) =\{X\mid |\Psi(z)\cup X|-|\Psi(z)|<m-c\}= \{X\mid | X \setminus \Psi(z)|<m-c\}$$
 be the set of bad $X$s for node $z$,  the parent of node $z$   will take the responsibility to remove all the bad $X$s against node $z$, then any $X$ in $\mathcal{V}_z$ is good for $z$ to choose.

Now we show how to construct the tree  recursively and lower bound the size of $\mathcal{H}$ which is exactly the number of leaves in the tree. Set $h=  2^{\lceil(1-\epsilon)n\rceil}$. Let $z$ be some node of $T(\mathcal{S})$ at depth $d\leq  h$, the node $z$ is associated with a subset $\mathcal{Z} \subseteq \mathcal{S}$. Initially, if $z$ is the root of $T(\mathcal{S}^\prime)$,   the trace $\Psi(z)$ at root $z$ is the  empty set,  the set of bad $X$s for $z$ is $\Phi(z) =\{X\mid | X \setminus \Psi(z)|<m-c\} = \{X\mid |X| <m-c\}$ where $|X|$ is the number of distinct rows in $X$, recall that we treat $X$ as a trace of its rows. Since root $z$ has no parent, we have to remove all bad $X$s for root $z$ in advance and $z$ is associated with a subset $\mathcal{S}^\prime \subseteq \mathcal{S}$ such that
\[\mathcal{S}^\prime =\left \{ (f,X)\mid (f,X) \in \mathcal{S}, X \not\in \{X \mid |X| < m-c\} \right\}.\]
Let $\Gamma_\mathcal{Z} = (\mathcal{U}_\mathcal{Z}, \mathcal{V}_\mathcal{Z},\mathcal{E}_\mathcal{Z})$ be the domain graph of $\mathcal{Z}$.  If $z$ is at depth $h$, then $z$ is a leaf, otherwise, we recursively construct a tree $T(\mathcal{Z})$ rooted at $z$  by attaching a set of  sub-trees to the node $z$.  Now since $z$'s parent has removed  all bad $X$s against $z$, all $X$s in  $\mathcal{V}_z$ are good. Let  $ X_z$ be some vertex of maximal degree in $\mathcal{V}_z$, then the  trace of each $z$'s children is $\Psi(z) \cup X_z$, now we want to remove all bad $X$s against $z$'s children, and the set of bad $X$ against $z$'s children is following set
\[\Phi^\prime(z) = \{X\mid |X \setminus (\Psi(z)\cup X_z)|<m-c\}.\]
After choosing $X_z$ and removing all bad $X$s against $z$'s children, let
\begin{align*}
\mathcal{Z}^\prime =\{(f, X) \mid (f,  X_z)\in \mathcal{Z}, 
(f, X)\in \mathcal{Z},
X \notin \Phi^\prime(z)  \}.
\end{align*}
$\mathcal{Z}^\prime$  is obtained from $\mathcal{Z}$ as follows. At first, remove all $f$s in $\mathcal{U}_\mathcal{Z}$ such that $(f,  X_z)$ is not in $\mathcal{Z}$, then for the remaining $f$s, remove every $(f,  X)$ such that $X\in \Phi^\prime(z)$ which is bad  against $z$'s children.

Now for every $a\in \{0,1\}^m$, let $\mathcal{Z}_{a}^\prime = \{(f, X) \mid (f, X) \in \mathcal{Z}^\prime ,f(X_z)=a\}$.
If $\mathcal{Z}_{a}^\prime
$ is not empty, there is a subtree $T(\mathcal{Z}_{a}^\prime)$ attached to the node $z$. Given two distinct subtrees $T(\mathcal{Z}^\prime_{a1}),T(\mathcal{Z}^\prime_{a_2})$, let $\Gamma_{\mathcal{Z}^\prime_{a_1}},\Gamma_{\mathcal{Z}^\prime_{a_2}}$ be  domain graphs of $\mathcal{Z}^\prime_{a_1},\mathcal{Z}^\prime_{a_2}$ respectively, then  $\mathcal{U}_{\mathcal{Z}^\prime_{a_1}}\cap \mathcal{U}_{\mathcal{Z}^\prime_{a_2}} =\emptyset$, since for every $f \in \mathcal{U}_{\mathcal{Z}^\prime_{a_1}}, g \in \mathcal{U}_{\mathcal{Z}^\prime_{a_2}}$, $f(X_z)=a_1 \neq a_2 =g(X_z)$. Thus recursively, for every two nodes $z_1,z_2$ at the same depth,  $ \mathcal{U}_{z_1}\cap \mathcal{U}_{z_2} =\emptyset$, and let $X$ be the label of the two nodes $z_1,z_2$' lowest common ancestor, for all $f\in  \mathcal{U}_{z_1}, g\in \mathcal{U}_{z_2}, f(X) \neq  g(X)$. Finally, for every two leaves  with $\mathcal{U}_{\ell_1}$ and $\mathcal{U}_{\ell_2}$, this is also true.

Now we are ready to lower bound the number of leaves in $T(\mathcal{S})$ by lower bounding the number of nodes at depth $d$.  The idea is to show the total number functions in these nodes is large and  the number of function in each single node is small. Since for every two nodes $z_1,z_2$ at the same depth,   $ \mathcal{U}_{z_1}\cap \mathcal{U}_{z_2} =\emptyset$, there must be many such nodes.

 Let $z$ be some node of the  tree $T(\mathcal{S})$ at depth $d\leq h$ labeled with $X_z$ corresponding to a root node of a subtree $T(\mathcal{Z})$ for some $\mathcal{Z}\subseteq \mathcal{S}.$ Let $\Gamma_\mathcal{Z} = (\mathcal{U}_\mathcal{Z}, \mathcal{V}_\mathcal{Z},\mathcal{E}_\mathcal{Z})$ be the domain graph of $\mathcal{Z}$. Let $T(\mathcal{Z}_{a_1} ), \ldots , T(\mathcal{Z}_{a_k} )$ be the subtrees attached to $z$ and $z_{a_1}, \ldots ,z_{a_k}$ be the roots of these subtrees respectively. Note that for every $i\in[k]$, trace $\Psi(z_{a_i})=\Psi(z) \cup X_z $ and   $\Phi(z_{a_i})=\Phi^\prime(z)=\{X\mid |X \setminus (\Psi(z)\cup X_z)|<m-c\}$.  Recall that $\mathcal{U}_{\mathcal{Z}_{a_i}}\cap \mathcal{U}_{\mathcal{Z}_{a_j}}=\emptyset$ for all $i\neq j$, let $\Gamma_{\mathcal{Z}^\prime }= (\mathcal{U}_{\mathcal{Z}^\prime }, \mathcal{V}_{\mathcal{Z}^\prime },\mathcal{E}_{\mathcal{Z}^\prime })$ be the domain graph of $\mathcal{Z}^\prime $, then $\mathcal{U}_{\mathcal{Z}_{a_1}} \cup \ldots \cup \mathcal{U}_{\mathcal{Z}_{a_k}} = \mathcal{U}_{\mathcal{Z}^\prime}$.
 Now let $$\mathcal{Z}^{\star} = \{(f, X) \mid (f,  X_z)\in \mathcal{Z}, (f, X)\in \mathcal{Z}\}$$ where $\mathcal{Z}^{\star}$  is obtained from $\mathcal{Z}$ by  collecting all $f$s in $\mathcal{U}_\mathcal{Z}$ such that $(f,  X_z)$ is  in $\mathcal{Z}$, then $\mathcal{U}_{\mathcal{Z}^{\star}} =  \mathcal{U}_{\mathcal{Z}^\prime}$. To see why this is true, we have to lower bound the degree of every $f \in \mathcal{U}_{\mathcal{Z}^\prime}$ in the domain graph $\Gamma_{\mathcal{Z}^\prime}$.

 Firstly, we  show for every node $z$ associated with some set $\mathcal{Z}\subseteq \mathcal{F}\times \mathcal{X}$, $\mathcal{X}_{\mathcal{Z},f} = \mathcal{X}_{\mathcal{S},f} \setminus \Phi(z)$ by induction on the depth of the node. Recall that when $z$ is the root node, $z$ is associated with $\mathcal{S}^\prime =\{ (f,X)\mid (f,X) \in \mathcal{S}, X \not\in \Phi(z)\}$, that is for every $f\in \mathcal{U_{S^\prime}}$, $\mathcal{X}_{\mathcal{S^\prime},f} = \mathcal{X}_{\mathcal{S},f} \setminus \Phi(z)$. Assume $z$ is node which is   associated with $\mathcal{Z}$, for every $f\in \mathcal{U}_\mathcal{Z}$,  $\mathcal{X}_{\mathcal{Z},f} = \mathcal{X}_{\mathcal{S},f} \setminus \Phi(z)$. Let $z_{a}$ be a child of $z$ and $z_{a}$ is associated with set $\mathcal{Z}_{a}$, then for every $f\in \mathcal{U}_{\mathcal{Z}_{a}}$, we have
  \begin{align*}
  \mathcal{X}_{\mathcal{Z}_{a},f}  & = \mathcal{X}_{\mathcal{Z},f}\setminus \Phi(z_a) \\ 
  & =(\mathcal{X}_{\mathcal{S},f} \setminus \Phi(z)) \setminus \Phi(z_a) \text{, by induction hypothesis}\\
  & =\mathcal{X}_{\mathcal{S},f} \setminus \Phi(z_a) \text{ , since $\Phi(z) \subseteq \Phi(z_a)$}
  \end{align*}
  as required.
  Now for every $f\in \mathcal{U}_\mathcal{Z}$,  we have $|\mathcal{X}_{\mathcal{Z},f}| \geq |\mathcal{X}_{\mathcal{S},f}| - |\Phi(z)|$. To proceed, we have to up bound $|\Phi(z)|$ as follows.
 \begin{align*}
 |\Phi(z)| &\leq |\{X\mid|X \setminus \Psi(z)|<m-c\}| 
  =\sum_{i=0}^{m-c-1}|\{X\mid |X \setminus \Psi(z)|=i\}| \\
 &\leq \sum_{i=0}^{m-c-1} \binom{2^n-|\Psi(z)|}{i} \cdot \binom{m}{i} \cdot (i+|\Psi(z)|)^{m-i} \\
  &\leq \sum_{i=0}^{m-c-1} 2^{ni} \cdot 2^m\cdot (m+md)^{m-i} , \text{since } |\Psi(z)| \leq md,i \leq m \\
&\leq \sum_{i=0}^{m-c-1} 2^{ni} \cdot 2^m \cdot 2^{((1-\epsilon)n+\log m +2)(m-i)} , \text{since } d \leq 2^{\lceil(1-\epsilon)n\rceil}\\
&= \sum_{i=0}^{m-c-1} 2^{mn} \cdot 2^{ (-\epsilon n+\log m +2)(m-i)} \cdot 2^m\\
&\leq  \sum_{i=0}^{m-c-1} 2^{mn} \cdot 2^{(-\epsilon n+\log m +2)(c+1)+ m}, \text{since} -\epsilon n+\log m +2<0, m-i\geq  c+1 \\
&\leq  2^{mn} \cdot 2^{(-\epsilon n+\log m +2)(c+1)+ m +\log m}\\ 
 &\leq 2^{-m} \cdot  2^{mn},  \text{since } c\geq \frac{2m+\log m}{\epsilon n-\log m-2}.
 \end{align*}

Thus, for every node $z$ associated with $\mathcal{Z}$,  for every $f\in \mathcal{U}_\mathcal{Z}$,  we have $|\mathcal{X}_{\mathcal{Z},f}| \geq 2^{t-m}\cdot 2^{mn}-  2^{-m} \cdot  2^{mn} \geq 2^{t-m-1}\cdot 2^{mn} \gg 0$. Now we show  $\mathcal{U}_{\mathcal{Z}^{\star}} =  \mathcal{U}_{\mathcal{Z}^\prime}$ 
 where $\mathcal{Z}^\prime$  is obtained from $\mathcal{Z}^{\star}$ by  removing bad $X$s, after the removal, for  every $f$ in $\mathcal{U}_{\mathcal{Z}^{\star}}$, $\mathcal{X}_{\mathcal{Z}^\prime,f}$ is still not empty  and $f$ remains  in $\mathcal{U}_{\mathcal{Z}^\prime}$. More formally,  for  every $f$ in $\mathcal{U}_{\mathcal{Z}^{\star}}$, $\mathcal{X}_{\mathcal{Z}^\prime,f}=\mathcal{X}_{\mathcal{Z}^\star,f}\setminus \Phi^\prime (z)=\mathcal{X}_{\mathcal{Z},f}\setminus \Phi^\prime (z)=\mathcal{X}_{\mathcal{S},f}\setminus \Phi^\prime (z)$. Let $z_a$ be some child of $z$, recall that $\Phi^\prime (z) =\Phi(z_a)$, thus for  every $f$ in $\mathcal{U}_{\mathcal{Z}^{\star}}$,  $\mathcal{X}_{\mathcal{Z}^\prime,f}=\mathcal{X}_{\mathcal{S},f}\setminus \Phi (z_a)$. Similarly,  $|\mathcal{X}_{\mathcal{Z}^\prime,f}| \geq |\mathcal{X}_{\mathcal{S},f}| - |\Phi(z_a)| \geq 2^{t-m-1}\cdot 2^{mn}\gg 0$ since  $|\Phi(z_a)|$ is also no larger than $2^{-m} \cdot  2^{mn}$. Particularly, we have $\mathcal{U}_{\mathcal{S}} =  \mathcal{U}_{\mathcal{S}^\prime}$. 

 Given that
$ X_z$ is a vertex of maximal degree in $\mathcal{V_Z}$ and $|\mathcal{V_Z} | \leq | \mathcal{X}| =  2^{mn}$,  the number of functions in the subtrees can be lower
bounded as follows
\begin{align*}
|\mathcal{U}_{Z_{a_1}} \cup \ldots \cup \mathcal{U}_{Z_{a_k}}|
&=|\mathcal{U}_{\mathcal{Z}^{\prime}}| =|\mathcal{U}_{\mathcal{Z}^{\star}}|
\geq \frac{|\mathcal{E_Z}|}{|\mathcal{V_Z}|} 
\geq  \frac{|\mathcal{U_Z}| \cdot  \min_{f \in\mathcal{U_Z} }|\mathcal{X}_{\mathcal{Z},f}|}{ 2^{mn}} \\
&\geq  \frac{|\mathcal{U_Z}| \cdot  2^{t-m-1}\cdot 2^{mn}}{ 2^{mn}} \\
&= \frac{|\mathcal{U_Z}|}{2^{m+1-t}}.
\end{align*}


Thus by induction the total number of functions that appear in the nodes at depth $d$ is at least
\[
 \frac{ |\mathcal{U_S}|}{2^{(m+1-t)d}}.
\]
Now we are ready to lower bound the number of nodes at some depth $d$. Let $z$ be a node, then for every $f\in \mathcal{U}_z, x \in \Psi(z)$, $f(x)$ is the same, so the number of distinct functions  in $\mathcal{U}_z$ is at most $2^{2^n}/2^{|\Psi(z)|} \leq 2^{2^n-(m-c)d}$. The number of nodes at depth $d$ is at least the total number of functions at depth $d$ divided by the upper bound on the number of functions in one node, that is
\begin{align*}
 \frac{ |\mathcal{U_S}|}{2^{(m+1-t)d}\cdot 2^{2^n-(m-c)d}}
=\frac{ 2^{(t-c-1)d} |\mathcal{U_S}|}{2^{2^n}}.
\end{align*}
Since  by assumption $|\mathcal{S}|\geq 2^{t-m}\cdot  |\mathcal{F}|\cdot |\mathcal{X}|$ and  $|\mathcal{F}|   \geq 2^{-2^{(1-\epsilon)n}} \cdot 2^{2^n} \geq 2^{-h} \cdot 2^{2^n}$, the size of $\mathcal{U_S}$ is at least $$\frac{|\mathcal{S}|}{|\mathcal{X}|}\geq \frac{2^{t-m}\cdot  |\mathcal{F}|\cdot |\mathcal{X}|}{|\mathcal{X}|} \geq 2^{t-m}\cdot 2^{-h}\cdot  2^{2^n},$$ the number of leaves at
depth $h=2^{\lceil (1-\epsilon)n\rceil}$ is at least
\begin{align*}
\frac{  2^{(t-c-1)h} \cdot 2^{t-m} \cdot 2^{-h} \cdot 2^{2^n}}{2^{2^n}}
&\geq  2^{(t-c-3)h}, \text{since } \epsilon <1-\frac{\log m}{n}, m\leq 2^{(1-\epsilon)n}\\
&\geq 2^h,\text {since } t\geq c+4\\
&=2^{2^{\lceil (1-\epsilon)n\rceil}}
\end{align*}
as required.
\end{proof}

\begin{thm}[The boosting theorem]\label{3.5}Let $m,n$ be integers  such that $m\geq 1,n> 2\log m+2$. Let $\epsilon \in ( \frac{\log m +2}{n},1-\frac{\log m}{n})$ be a parameter and $c,t,s$ be integers satisfying $c\geq \frac{2m+\log m}{\epsilon n-\log m -2},t\geq c+4 ,s \leq m-t-1$. Let  $\mathcal{F}$ be a set of functions $f:\{0,1\}^n \rightarrow \{0,1\}$ such that $|\mathcal{F}|\geq 2^{-2^{(1-\epsilon)n}} \cdot 2^{2^n}$, let $\mathcal{X}$ be the set $\{0,1\}^{m\times n}$, if $\mathsf{CC^{phd}}(\textnormal{U}_m \diamond \textnormal{MUX}_\mathcal{F}) \geq s$,  $\mathsf{CC^{phd}}(\textnormal{U}_m \diamond \textnormal{MUX}_\mathcal{F}) \geq s+(1-\epsilon) n-\log m-\log n-6.$
\end{thm}
\begin{proof}Given any partially half-duplex protocol $\Pi$ for $\textnormal{U}_m \diamond \textnormal{MUX}_\mathcal{F}$, let $d$ be the depth of protocol $\Pi$, since $\mathsf{CC^{phd}}(\textnormal{U}_m \diamond \textnormal{MUX}_\mathcal{F}) \geq s$, $d\geq s$. Let $\mathcal{D}=\{((f,X),(f,X))\mid f\in \mathcal{F},X\in \mathcal{X}\}$, by Fact \ref{2.29}, there is a transcript $\tau\in \{0,1\}^s$ and a subset of inputs  $\mathcal{D}^\prime \subseteq \mathcal{D}$ such that every input in $\mathcal{D}^\prime$ is consistent with $\tau$ and   $|\mathcal{D}^\prime|  \geq |\mathcal{D}|/2^s$, let $\mathcal{T}=\{(f,X)\mid((f,X),(f,X)) \in \mathcal{D}^\prime\}$ then $|\mathcal{T}| \geq 2^{-s} \cdot |\mathcal{F}|\cdot |\mathcal{X}| \geq 2^{t+1-m} \cdot |\mathcal{F}|\cdot |\mathcal{X}|.$ Removing every $f$ such that $|\mathcal{X}_{\mathcal{T},f}| <2^{t-m} \cdot |\mathcal{X}|$ in $\mathcal{T}$, and obtain  $\mathcal{S} =\{(f,X) \mid (f,X) \in \mathcal{T}, |\mathcal{X}_{\mathcal{T},f}| \geq  2^{t-m} \cdot |\mathcal{X}|\}$, then $|\mathcal{S}| \geq |\mathcal{T}|-|\mathcal{F}| \cdot 2^{t-m} \cdot |\mathcal{X}|\geq  2^{t-m} \cdot|\mathcal{F}| \cdot |\mathcal{X}|$, and let $\Gamma_\mathcal{S}= (\mathcal{U}_\mathcal{S}, \mathcal{V}_\mathcal{S},\mathcal{E}_\mathcal{S})$ be the domain graph of $\mathcal{S}$,  for every $f\in \mathcal{U}_\mathcal{S}$,   $|\mathcal{X}_{\mathcal{S},f}| \geq 2^{t-m}\cdot |\mathcal{X}|$. 
	
Apply Lemma \ref{3.4} with $\mathcal{S}$ and parameters $m,n,\epsilon,c,t$, then there is a set $\mathcal{H}\subseteq \mathcal{U}_\mathcal{S}$ of size at least $2^{2^{(1-\epsilon)n}}$  such that for all distinct $f, g \in \mathcal{H}$, there exists an $ X:(f,  X), (g,  X)\in \mathcal{S}$, and $f(X) \neq  g(X)$. Apply Lemma \ref{3.2} with the transcript $\tau$ and the set $\mathcal{H}$, we have $d\geq s+(1-\epsilon) n-\log m-\log n-6$ as required.
\end{proof}

Now we prove Theorem \ref{1.6} rephrased as follows.
\begin{thm}\label{3.6}Let $m,n$ be integers  such that $m\geq 1,n> 2\log m + \log n+9$. Let $\epsilon \in ( \frac{\log m +2}{n},1-\frac{\log m+\log n +7}{n})$. Let  $\mathcal{F}$ be a set of functions $f:\{0,1\}^n \rightarrow \{0,1\}$ such that $|\mathcal{F}|\geq 2^{-2^{(1-\epsilon)n}}\cdot 2^{2^n}$, then $$\mathsf{CC^{phd}}(\textnormal{U}_m \diamond \textnormal{MUX}_\mathcal{F}) \geq m+n-\left\lceil \frac{2m+\log m}{\epsilon n-\log m -2} \right\rceil-\epsilon n-\log m-\log n-11.$$
Furthermore, let  $m=\omega(\log^2 n),n =\omega(\sqrt{m})$,  $\epsilon =\frac{\sqrt{m}}{n}$, we have
$$\mathsf{CC^{phd}}(\textnormal{U}_m \diamond \textnormal{MUX}_\mathcal{F}) \geq m+n-\left\lceil \frac{2m+\log m}{\sqrt{m}-\log m -2} \right\rceil-\sqrt{m}-\log m -\log n-11=m+n-O(\sqrt{m}).$$
\end{thm}
\begin{proof}In the beginning, set $c= \lceil \frac{2m+\log m}{\epsilon n-\log m -2} \rceil,t= c+4,s=0$ where $s$ is current known lower bound of $\mathsf{CC^{phd}}(\textnormal{U}_m \diamond \textnormal{MUX}_\mathcal{F})$. Repeatedly applying Theorem \ref{3.5}, after each application, we have $s\leftarrow s+(1-\epsilon) n-\log m-\log n-6$. Since $\epsilon \in ( \frac{\log m +2}{n},1-\frac{\log m+\log n +7}{n})$, $(1-\epsilon)n -\log m-\log n -6 \geq 1$, that is every application will increase the complexity at least one. This repetition will not end until $s=m-t-1$, when $s=m-t-1$, apply  Theorem \ref{3.5} for the last time, and obtain 
	\begin{align*}
	s &\geq m-t-1+(1-\epsilon) n-\log m -\log n-6 \\
	&=m-\left\lceil \frac{2m+\log m}{\epsilon n-\log m -2} \right\rceil-4-1 +(1-\epsilon) n -\log m-\log n -6\\
	&=m+n-\left\lceil \frac{2m+\log m}{\epsilon n-\log m -2} \right\rceil-\epsilon n-\log m-\log n -11
	\end{align*}
	as required.
\end{proof}

\section{A  Composition Theorem of a Universal Relation and Most  Functions}\label{s4}
In this section we prove when $m,n$ are in proper range, for most functions $f:\{0,1\}^n \rightarrow \{0,1\}$, the communication complexity of $\textnormal{U}_m \diamond\textnormal{KW}_f$ is at least $m+n-O(\sqrt{m})$. At first, we need the following lemma which transforms the complexity of $\textnormal{U}_m \diamond \textnormal{MUX}_\mathcal{F}$ in the partially half-duplex model to the complexity of $\textnormal{U}_m \diamond\textnormal{KW}_f$ for some function $f\in \mathcal{F}$ in the standard model of communication. The lemma is proved with the same idea in \cite{DBLP:conf/coco/MihajlinS21}.
\begin{lem}\label{4.1}Let  $\mathcal{F}$ be a set of functions $f : \{0,1\}^n \rightarrow \{0,1\}$, then  $$\max_{f\in  \mathcal{F}}\mathsf{CC}(\textnormal{U}_m \diamond\textnormal{KW}_f) 
	\geq  \mathsf{CC^{phd}}(\textnormal{U}_m \diamond \textnormal{MUX}_\mathcal{F})-\log mn-2.$$
\end{lem}
\begin{proof}Let $ d= \max_{f\in  \mathcal{F}}\mathsf{CC}(\textnormal{U}_m \diamond\textnormal{KW}_f)$. For every $f\in \mathcal{F}$,  Alice and Bob hold the same optimal standard protocol $\Pi_f$ which depth is no larger than $d$. Now we can leverage this to construct a partially half-duplex protocol for $\textnormal{U}_m \diamond \textnormal{MUX}_\mathcal{F}$. Given input $(f,X)$, Alice simulates the protocol $\Pi_f$ on input $X$. Similarly, given input $(g,Y)$, Bob simulates the protocol $\Pi_g$ on $Y$. When Alice performs $t$ rounds and reaches some leaf labeled with $(i,j)$ in protocol $\Pi_f$, if $t < d$, Alice performs another $d-t$ round of sending $1$. Similarly,  when Bob performs $t^\prime$ rounds and reaches some leaf labeled with $(i^\prime,j^\prime)$ in protocol $\Pi_g$, if $t^\prime < d$, Bob performs another $d-t^\prime$ round of receiving.\footnote{In the  proof of a similar lemma in \cite{DBLP:conf/coco/MihajlinS21} by Ivan Mihajlin and Alexander Smal, they ask both Alice and Bob to perform the action of receiving after reaching leaves, this is problematic, when Alice and Bob are given the same function, they also perform non-classical rounds after reaching leaves.} After both players spend exact $d$ rounds, they start to verify  that Alice's answer is correct. Alice sends $(i,j)$ and $X_{i,j}$ to Bob,  Bob replies with $1$ if  $X_{i,j}  \neq Y_{i,j}$ and $0$ otherwise. Finally, they output $(i,j)$ if Alice's answer is correct and $\bot$ otherwise.
	
	When Alice and Bob are given the same function $f$, they must perform $t$ classical rounds and  reach the same leaf in protocol tree $\Pi_f$ since they simulate the same protocol $\Pi_f$. In the next $d-t$ round, Alice sends $1$ and Bob receives $1$, after that they perform classical rounds to verify Alice's answer is correct. Thus, above protocol is indeed a correct partially half-duplex protocol and it spends $d+\log mn +2$ bits communication. That is $\mathsf{CC^{phd}}(\textnormal{U}_m \diamond \textnormal{MUX}_\mathcal{F}) \leq d+\log mn +2$ as required.
\end{proof}
Now we prove Theorem \ref{1.5} rephrased as follows.
\begin{thm}Let  $m=\omega(\log^2 n),n =\omega(\sqrt{m}),\epsilon =\frac{\sqrt{m}}{n}$, there are at least $2^{2^n}(1-2^{-2^{(1-\epsilon)n}})$ distinct functions $f:\{0,1\}^n \rightarrow \{0,1\}$ such that  $\mathsf{CC}(\textnormal{U}_m \diamond \textnormal{KW}_f) \geq m+n-O(\sqrt{m})$.
\end{thm}
\begin{proof}In the beginning, let $\mathcal{F}$ be the set of all functions $f:\{0,1\}^n \rightarrow \{0,1\}$ and $|\mathcal{F}|=2^{2^n}$, apply Theorem \ref{3.6} with $\mathcal{F}$, we have $\mathsf{CC^{phd}}(\textnormal{U}_m \diamond \textnormal{MUX}_\mathcal{F}) \geq m+n-O(\sqrt{m}).$ By Lemma~\ref{4.1}, there exists a function $f$ such that $\mathsf{CC}(\textnormal{U}_m \diamond \textnormal{KW}_f) \geq m+n-O(\sqrt{m})-\log mn-2 =m+n-O(\sqrt{m})$. Now set $\mathcal{F}\leftarrow \mathcal{F} \setminus \{f\}$, repeat this process until $|\mathcal{F}| < 2^{-2^{(1-\epsilon)n}} \cdot 2^{2^n}$, then we have found at least $2^{2^n}-2^{-2^{(1-\epsilon)n}} \cdot 2^{2^n}$ functions $f:\{0,1\}^n \rightarrow \{0,1\}$ such that  $\mathsf{CC}(\textnormal{U}_m \diamond \textnormal{KW}_f) \geq m+n-O(\sqrt{m})$.
\end{proof}
\section{Conclusion and Discussion}\label{s5}
Here we make some discussion about our results and point out some future directions. As mentioned before, our method can be used to obtain a similar result for function bundles. That is for  most $(m,n)$  function bundles $F$, $\mathsf{CC}(\textnormal{KW}_F)$ is about $m+n-O(\sqrt{m})$, this can be done with a slightly different way of restriction. We also note that our method can be applied  to other related conjectures in \cite{DBLP:journals/cc/DinurM18}. Take  Conjecture 9.4  in \cite{DBLP:journals/cc/DinurM18} as example, Dinur and Meir conjectured given a subset $\mathcal{X} \subseteq \{0, 1\}^{m \times n}$ with  density at least $2^{-(m-\tilde{O} (\sqrt{m}))}$, then the restriction of $\textnormal{U}_m \diamond \textnormal{KW}_f$ to  $\mathcal{X} \times \mathcal{X}$ has communication complexity at least $\mathsf{CC}(\textnormal{KW}_f)-\tilde{O} (\sqrt{m})$. With our method, we can show that when choosing $f,\mathcal{X}$ randomly, it is true with high probability. But it is not clear whether our method is helpful in the case of  $1$-out-of-$k$ problem of KW relation \cite{DBLP:journals/cc/DinurM18}. Furthermore, comparing to the optimal lower bound in the case of $\tKW_{f}\diamond \operatorname{U}_{n}$, there still is room for improvement, thus the  question is  can we prove a lower bound for $\textnormal{U}_m \diamond \textnormal{KW}_f$ with poly-logarithmic additive loss. We also suspect that our result can be extended to   a slightly weaker lower bound in terms of protocol size like those in \cite{DBLP:journals/siamcomp/GavinskyMWW17,Koroth2018}, but we haven't fully verify it.

The next major step is to consider the composition of two multiplexors. Let $\mathcal{F}$ be the set of all functions $f:\{0,1\}^m \rightarrow \{0,1\}$, $\mathcal{G}$ be  the  set of all functions $g:\{0,1\}^n \rightarrow \{0,1\}$ and $\Delta = \mathcal{F} \times \mathcal{G}$. In  KW relation $\textnormal{MUX}_m \diamond \textnormal{MUX}_n$, Alice gets a pair of functions $(f,g)\in \Delta$ and a Boolean matrix $X\in \{0,1\}^{m \times n}$,  Bob gets a pair of functions $(f^\prime,g^\prime)\in \Delta$ and a Boolean matrix $Y\in \{0,1\}^{m \times n}$,  their goal is to find an entry $(i, j)$ such that $X_{i, j} \neq Y_{i, j}$. If $(f,g) \neq (f^\prime,g^\prime)$ or $f \diamond g(X)=f^\prime \diamond g^\prime(Y)$, they can also output $\bot$.  We think it may be easier to prove lower bound for composition of two multiplexors than composition of a function and a multiplexor. Comparing to the KW relation of a function, the multiplexor looks more like the universal relation. 
But our current way of restriction  won't immediately work in the case of two multiplexors. When constructing the binary tree for the second stage,  in each step downward, the number of functions in each child decreases by  a factor of (at most) $2$ while the total number of functions in all its children decreases by a  (average) factor of approximate $2^{m}$. Due to the fact that the composite function $f\diamond g$ takes Boolean values, the protocol can easily divide a set $\mathcal{S} \subseteq \Delta \times \mathcal{X}$ into two parts such that in each part $f\diamond g(X)$ is the same for every $(f\diamond g,X)\in \mathcal{S}$. Thus considering a square $\mathcal{S}\times \mathcal{S}$ is not helpful anymore in this case, we should consider general rectangle like those in  \cite{DBLP:journals/cc/Meir20,DBLP:journals/eccc/Meir23} and  new ideas are needed. Maybe we should try our method in the case of strong composition of two multiplexors in the first place. A less ambitious question is to show a composition theorem of a parity function and a multiplexor.

\section*{Acknowledgments}
The author is  grateful to Ivan Mihajlin and Alexander Smal for many detailed comments and valuable suggestions that greatly improved the presentation of this paper.

\appendix

\bibliographystyle{alpha}
\bibliography{hw}

\end{document}